\documentclass{aa}
\usepackage{natbib}
\usepackage{graphicx}
\usepackage{txfonts}
\usepackage{booktabs}
\usepackage[colorlinks=true,citecolor=blue,linkcolor=magenta,urlcolor=blue]{hyperref}
\usepackage{caption}
\usepackage{multirow}%
\usepackage{bigdelim}%
\DeclareCaptionLabelFormat{continued}{#1~#2 (continued)}%
\newcommand{\msun}{$M_{\odot}$}

\newcommand{\teff}{$T_\mathrm{eff}$}
\newcommand{\logg}{$\log g$}
\newcommand{\logy}{$\log n(\mathrm{He}) / n(\mathrm{H})$}
\newcommand{\Z}{[Fe/H]}
\newcommand{\alp}{[$\alpha/\mathrm{Fe}$]}
\newcommand{\lsiv}{\object{LS\,IV$-$14$\degr$116}}
\newcommand{\feige}{\object{Feige\,46}}
\newcommand{\EC}{EC\,22536$-$5304}
\def\BD+25{BD+25$^{\circ}\,$4655}
\newcommand{\tlusty}{\textsc{Tlusty}}
\newcommand{\synspec}{\textsc{Synspec}}
\newcommand{\synthe}{\textsc{Synthe}}
\def\A9{\textsc{Atlas}{\footnotesize9}}
\def\A12{\textsc{Atlas}{\footnotesize12}}
\newcommand{\kms}{km\,s$^{-1}$}

\newcommand{\uh}[1]{\textcolor{black}{#1}}
\newcommand{\todo}[1]{\textcolor{black}{#1}}
\newcommand{\up}[1]{\textcolor{black}{#1}}
\newcommand{\led}[1]{\textcolor{black}{#1}}

\begin{document}

\title{EC\,22536$-$5304: A lead-rich and metal-poor long-period binary}

\author{M.\ Dorsch\inst{\uh{1,2}}, C.\ S.\ Jeffery\inst{3}, \uh{A.\ Irrgang\inst{\uh{1}}, V.\ Woolf\inst{4},} U.\ Heber\inst{\uh{1}}}  

\institute{
Dr.\ Karl Remeis-Observatory \& ECAP, Friedrich-Alexander University Erlangen-N\"{u}rnberg,
Sternwartstr.\ 7, 96049 Bamberg, Germany
\email{matti.dorsch@fau.de} 
\and
Institut für Physik und Astronomie, Universität Potsdam, Haus 28, Karl-Liebknecht-Str.\ 24/25, 14476 Potsdam-Golm, Germany
\and
Armagh Observatory and Planetarium, College Hill, Armagh BT61 9DG, Northern Ireland
\and
Department of Physics, University of Nebraska at Omaha, 6001 Dodge St, Omaha, NE 68182-0266, USA
}

\date{Received ; accepted }

\abstract
{
Helium-burning hot \led{subdwarf} stars of spectral types O and B (sdO/B) are thought to be produced through various types of binary interactions. 
The helium-rich hot \led{subdwarf} star \EC\ was recently found to be extremely enriched in lead.
Here, we show that \EC\ is a binary star with a metal-poor \led{subdwarf} F-type (sdF) companion. 
We performed a detailed analysis of high-resolution SALT/HRS and VLT/UVES spectra, deriving metal abundances for the hot \led{subdwarf}, as well as atmospheric parameters for both components.
Because we consider the contribution of the sdF star, the derived lead abundance for the sdOB, $+6.3\pm 0.3$\,dex relative to solar, is even higher than previously thought. 
We derive \teff\,=\,$6210\pm70$\,K, \logg\,=\,$4.64\pm0.10$, \Z\,=\,$-1.95\pm0.04$, and \alp\,=\,$+0.40\pm0.04$ for the sdF component. 
Radial velocity variations, although poorly sampled at present, indicate that the binary system has a long orbital period of about 457 days.
This suggests that the system was likely formed through stable \up{Roche lobe} overflow (RLOF). 
A kinematic analysis shows that \EC\ is on an eccentric orbit around the Galactic centre.
This, as well as the low metallicity and strong alpha enhancement of the sdF-type companion, indicate that \EC\ is part of the Galactic halo or metal-weak thick disc.
As the first long-period hot \led{subdwarf} binary at \Z\,\led{$\lesssim$}\,$-1$, \EC\ may help to constrain the RLOF mechanism for mass transfer from low-mass, low-metallicity red giant branch (RGB) stars to main-sequence companions. 
}


\authorrunning{Dorsch et al.}
\titlerunning{EC\,22536$-$5304}

\maketitle

\section{Introduction}
Hot \led{subdwarf} stars form a heterogeneous class of evolved stars located at the \up{hot end of the horizontal branch} or beyond \citep[for reviews, see][]{heber09,heber16}. 
This implies that they burn helium either in the core or in a shell, while their hydrogen envelope is very thin or almost absent.
Accordingly, their masses are close to half that of the Sun.
Most B-type \led{subdwarf} stars (sdB) have helium-poor atmospheres due to atmospheric diffusion \citep{Hu2011,Michaud2011}. 
In contrast, the hotter sdO stars are often found to be extremely enriched in helium. 
There is also an intermediate class of hot \led{subdwarf} stars, termed intermediate helium sdOB (iHe-sdOB).
Their effective temperature and helium abundances are intermediate between the cooler helium poor sdBs and those of the hotter helium-rich sdO stars \citep[e.g.][]{nemeth12,Jeffery2021}. 
Recently, several members of this small population have been found to be extremely enriched in heavy elements such as strontium, yttrium, zirconium, or lead \citep[e.g.][]{naslim11,Jeffery2017,Dorsch2020}. 
The hot component of \EC\ was identified as an extremely lead-rich iHe-sdOB by \cite{Jeffery2019}; in fact, it is the most lead-rich star known to date.
Like the depletion of helium observed in the photospheres of most sdB stars, the strong enrichment in heavy metals is usually attributed to diffusion and selective radiative acceleration. 
However, quantitative predictions are still lacking for atomic diffusion in the atmospheres of iHe-sdOB stars.

A close inspection of new high-resolution spectra taken with SALT/HRS reveals a second component in the spectrum of \EC: 
a metal-poor, \led{subdwarf}-F-type (sdF) main-sequence (MS) star, which was not considered in the previous analysis. 
Many helium-poor sdB stars are found in binary systems with low-mass MS stars or white dwarfs at short orbital periods of the order of ten days or fewer \citep{Kupfer2015}.
They are thought to be formed following a common envelope (CE) phase, in which the red giant progenitor to the sdB has lost most of its hydrogen-rich envelope, just before it ignites helium burning in the core \citep{han02}. 

The orbital properties of all 23 solved helium-poor sdB stars in long-period binaries have recently been published by \cite{Vos2019}. 
They find that the orbital periods of these sdB + F/G/K-type systems range from about 500 to 1400 days. 
The hot \led{subdwarf} stars in such systems are the result of a stable Roche overflow (RLOF) as the progenitor star to the sdB reaches the tip of the red giant branch \citep[RGB,][]{Han2003,Chen2013}.
Only two helium-rich \led{subdwarf} stars in long-period binary systems are known:
the post-asymptotic giant branch (AGB) He-sdO HD\,128220\,B \citep{Rauch1993} and the He-sdO  HD\,113001\,B, a visually resolved binary with a very long orbital period that has likely not undergone mass transfer \citep{Tomley1970,Goy1980,Orlov2010}.
\led{Very recently, \cite{Nemeth2021} have found the He-poor sdOB component of the long-period binary system SB\,744 to be extremely enhanced in lead.}

\EC\ is the first \led{helium-rich} heavy-metal \led{subdwarf} found to be in a binary system (long- or short-period). 
As we show in the following, \EC\ is likely to be a long-period binary, and the first such system found at a metallicity  below \led{about} $\mathrm{[Fe/H]}=-1$. 
It therefore presents a unique opportunity to study the RLOF evolutionary scenario\todo{, especially once additional spectra become available that will further constrain} the orbital parameters of the system. 

In this paper, we present a detailed analysis of high-resolution spectra of \EC. 
The presence of the cool component is confirmed by an analysis of the spectral energy distribution (SED), which is described in Sect.~\ref{sect:sed}.
The available spectroscopic data are summarised in Sect.~\ref{sect:obs}.
Results from the SED and spectroscopic analysis (Sect.~\ref{sect:atm}) are combined with the parallax measurement provided by the \textit{Gaia} mission to derive stellar masses, radii, and luminosities in Sect.~\ref{sect:mass}. 
Because the cool companion (hereafter \EC\,B) contributes significantly in the optical range, we updated the metal abundances derived for the hot component (hereafter \EC\,A). 
A detailed account of the metal abundance analysis for the sdOB is given in Sect.~\ref{sect:metal}.
The radial velocities derived from the currently available spectra are used to perform \uh{a kinematic} analysis in Sect.~\ref{sect:velocity}.
The Galactic orbit of the system is characterised in Sect.~\ref{sect:kinematics}.
We summarise our results in Sect.~\ref{sect:conclusions}.

\section{Spectral energy distribution}\label{sect:sed}

\begin{figure}
    \vspace{0pt}
    \includegraphics[width=\columnwidth]{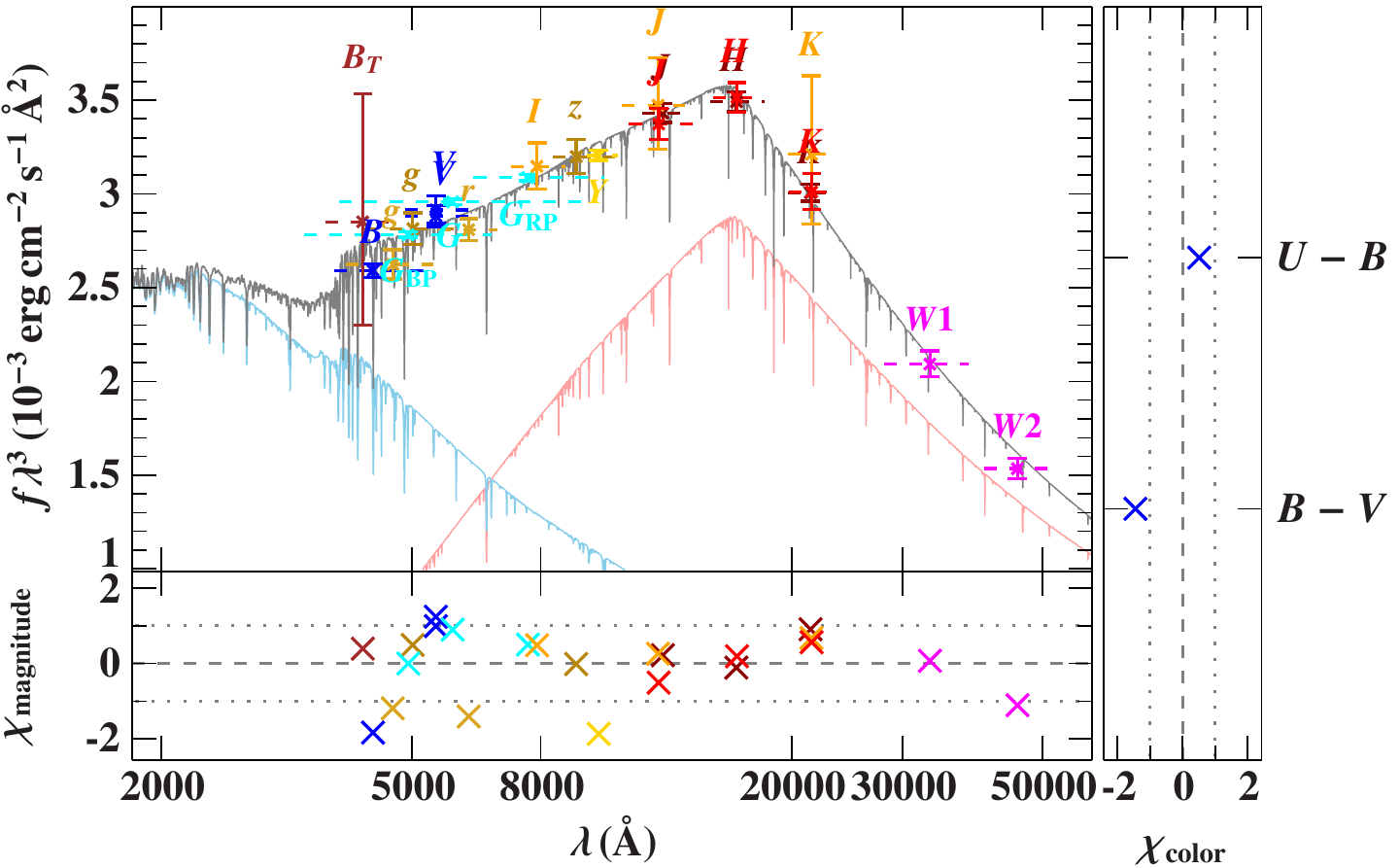}
    \vspace{0pt}
\caption{Photometric fit for \EC. Filter-averaged fluxes are shown as coloured data points that were converted from observed magnitudes.
Filter widths are indicated by dashed horizontal lines.
The grey line visualises the combined model spectrum while individual contributions are shown in blue (A) and red (B). 
The residual panels on the bottom and right side respectively show the differences between synthetic and observed magnitudes and colours.
The following colour codes are used to identify the photometric systems: 
Tycho \cite[brown,][]{Hog2000},
Johnson-Cousins \cite[blue,][]{Henden_2015,Kilkenny2016},
SDSS \cite[yellow,][]{Henden_2015}, 
SkyMapper \cite[dark yellow,][]{Wolf2018}, 
\textit{Gaia}  \cite[cyan,][]{Riello2020}, 
DENIS \cite[orange,][]{vizier:B/denis}, 
DES \cite[bright yellow,][]{Abbott2018}, 
VISTA \cite[dark red,][]{McMahon2013},
2MASS \cite[bright red,][]{Cutri2003_2MASS},
and WISE \cite[magenta,][]{Cutri2014}.
}
    \label{fig:EC:SED}
\end{figure}

\begin{table}
\centering
\caption{SED fit and spectroscopic fit results for the hot (A) and cool (B) components of \EC.}
\label{tab:EC:param}
\setstretch{1.15}
\begin{tabular}{lcc}
\toprule
\toprule
{} & SED fit & Spectral fit   \\
\midrule
$\log \Theta_\mathrm{A}\,\mathrm{(rad)}$ & $-11.09^{+0.09}_{-0.05}$ & -- \\
\midrule
$T_\mathrm{eff,A}$ (K) &  $38000^{+5000}_{-7000}$  & $38000 \pm 400$ \\
$\log g_\mathrm{A}$ &    & $\phantom{+}5.81 \pm 0.04$  \\
$\log n(\mathrm{He}) / n(\mathrm{H})$ &  & $-0.15 \pm 0.04$  \\
$v_\mathrm{tb,A}$ (km\,s$^{-1}$) &   & $\phantom{+}2.1\pm 0.2$\\
$v_\mathrm{rot}\sin i~_\mathrm{A}$ (km\,s$^{-1}$) & -- & $0.0^{+1.0}_{-0.0}$  \\
\midrule
$T_\mathrm{eff,B}$ (K)  &  $6460^{+90}_{-190}$  & $6210 \pm 70$ \\
$\log g_\mathrm{B}$  &    & $\phantom{+}4.64\pm 0.10$  \\
\Z$_\mathrm{B}$ &   & $-1.95 \pm 0.04$  \\
\alp$_\mathrm{B}$ &    & $+0.40 \pm 0.04$ \\
$v_\mathrm{tb,B}$ (km\,s$^{-1}$)  &  & $\phantom{+}1.83\pm 0.05$ \\
$v_\mathrm{rot}\sin i~_\mathrm{B}$ (km\,s$^{-1}$) & -- & $15.3\pm 0.2$  \\
\midrule
Surface ratio & $34\pm 9$ & $34 \pm 5$ \\
\bottomrule
\end{tabular}
\end{table}

Due to the large difference in their effective temperatures, both components of \EC\ are easily distinguished in the SED of the system. 
To obtain an initial estimate for the atmospheric parameters of both components, a first photometric fit was performed before the spectral analysis.
We obtained apparent magnitudes from the ultraviolet to the infrared to construct the SED of \EC.
A general description of the $\chi^2$ minimisation method used for the SED fit is given by \cite{Heber2018}. 
In short, model spectra of both components were converted to filter-averaged magnitudes and scaled to match the observed magnitudes. 
Free fit parameters were the effective temperatures of both components, the angular diameter of the sdOB $\Theta_\mathrm{A}$, and the surface ratio $A_\mathrm{sdF}/A_\mathrm{sdOB}$. %
Here, we used the same large model grids as for the spectroscopic analysis, which are described in Sect.~\ref{sect:atm}.
Interstellar reddening was considered as by \cite{Fitzpatrick2019}, assuming an extinction parameter of $R (55) = 3.02$. 
Since no reliable UV magnitudes are available, we constrained the colour excess not to exceed the line-of-sight value given by \cite{Schlegel1998}, \uh{$E(B-V)=0.126$\,mag}, which is reached in the fit.
The first fit was later refined by fixing the helium abundance of the hot component, the metallicity \up{and alpha enhancement} of the cool component, and both surface gravities to values derived from the spectral analysis.
The final SED fit is shown in Fig.~\ref{fig:EC:SED}. 
Table \ref{tab:EC:param} compares results of the SED fit with parameters derived from the spectral fit (see Sect.~\ref{sect:atm}).

Several heavy metal \led{subdwarf}s have been found to show peculiar low-amplitude photometric variability: the zirconium-rich \lsiv, \feige, and \uh{PHL\,417} \citep{Ahmad2005,Latour2019a,Ostensen2020}, as well as the lead-rich  UVO\,0825+15 \citep{Jeffery2017}. 
It would therefore be interesting to search for photometric variability in \EC\ as well.

\section{Spectroscopic observations}
\label{sect:obs}
\begin{table}
\setstretch{1.2}
\captionof{table}{Spectroscopic data available for \EC\tablefootmark{a}.}
\label{tab:obs}
\vspace{-11pt}
\begin{center}
\begin{tabular}{l c c c c c}
\toprule
\toprule
Spectrograph & Range / \AA  & $R$  & $n_\mathrm{exp}$ & S/N \\ 
\midrule
HRS/blue  &  $3895-5520$   &  43000  &   25  & 200 \\
HRS/red   &  $5500-8870$   &  41000  &   \phantom{0}9 & 120 \\
RSS   &  $3850-5100$   &  1.6\,\AA\tablefootmark{b}  &   \phantom{0}2 & 100 \\
UVES/blue  &  $3740-4525$   &  41000  &   \phantom{0}3  & \phantom{0}50 \\
UVES/red   &  $5655-9463$   &  42000  &   \phantom{0}2 & \phantom{0}25 \\
\bottomrule 
\end{tabular}
\end{center}
\vspace{-7pt}
\tablefoot{
\tablefoottext{a}{The signal-to-noise ratio is the maximum reached in a combined spectrum.}
\tablefoottext{b}{The resolution for RSS is given as $\Delta \lambda$.}
}
\end{table}

The spectroscopic data available for \EC\ are summarised in Table \ref{tab:obs}.
\EC\ has been observed extensively with the Southern African Large telescope (SALT) using both the Robert Stobie Spectrograph (RSS) and the High Resolution Spectrograph (HRS). 
The setups used for both spectrographs are described in detail in \cite{Jeffery2019}.
They used two long-slit RSS spectra taken in June 2018, as well as HRS spectra taken on 2017 May 18 and 2018 November 15, with exposure times of $2\times 2000$\,s on both occasions.
These individual HRS spectra have mean signal-to-noise ratios (S/N) of about 30.
Twenty-one additional blue HRS spectra were taken between 2019 May 9 and 23, increasing the combined S/N to about 200.
The HRS spectra consist of short \'echelle orders, which make order merging difficult.
We used a technique that normalises all orders simultaneously, ensuring continuity across the order overlaps, before stitching the individual orders together. 
Some residual anomalies persist, so we re-normalised the order-merged spectra before performing the spectral analysis. 
As a result, broad hydrogen and helium lines in these spectra could not be used to estimate atmospheric parameters.
The HRS spectra remained essential for the detection of weak metal lines of both components, as well as the \ion{Mg}{i} triplet.
Coverage of the \ion{Mg}{i} triplet is especially important because it is sensitive to the metallicity, alpha-enhancement, and surface gravity of the F-type companion. 

In addition to the blue HRS spectra, nine red HRS spectra of sufficient quality are available. 
These spectra were taken on the same dates as the blue HRS spectra. %
Their near infrared coverage is especially useful since it includes the \ion{Ca}{ii} triplet of the F-type star, as well as \ion{He}{i} 6678, 7065, and 7281\,\AA\ for the sdOB, all of which help to constrain the surface ratio.

We also make use of archival UVES spectra, which were obtained in October and December 2011 by M.~R.~Schreiber under Programme ID 088.D-0364(A).
Because normalisation issues are less pronounced in the blue UVES spectra than in the HRS spectra, UVES spectra proved to be valuable for the determination of atmospheric parameters from broad hydrogen and helium lines, despite their lower S/N.

\section{Spectral analysis and atmospheric parameters}
\label{sect:atm}

\begin{figure*}%
\centering
\includegraphics[width=0.99\textwidth]{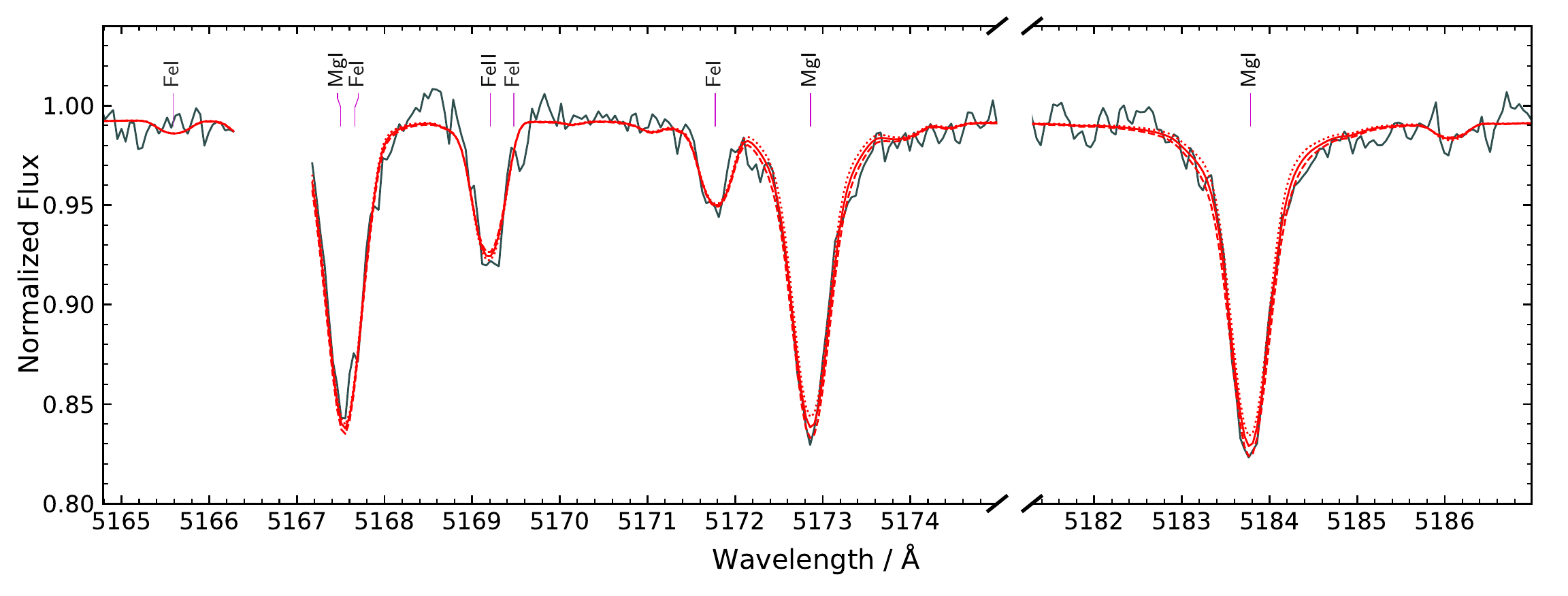}
\caption{ \ion{Mg}{i} triplet in the HRS spectrum of \EC\ (grey). 
The combined model spectrum (red) is the sum of the contributions of the sdOB and F-type star at the best fit.
The dashed spectrum was computed at $\log g_2$ = 4.9, while the dotted spectrum uses $\log g_2$ = 4.55.}
    \label{fig:EC:MgI}
\end{figure*}
\begin{figure*}%
\centering
\includegraphics[width=0.99\textwidth]{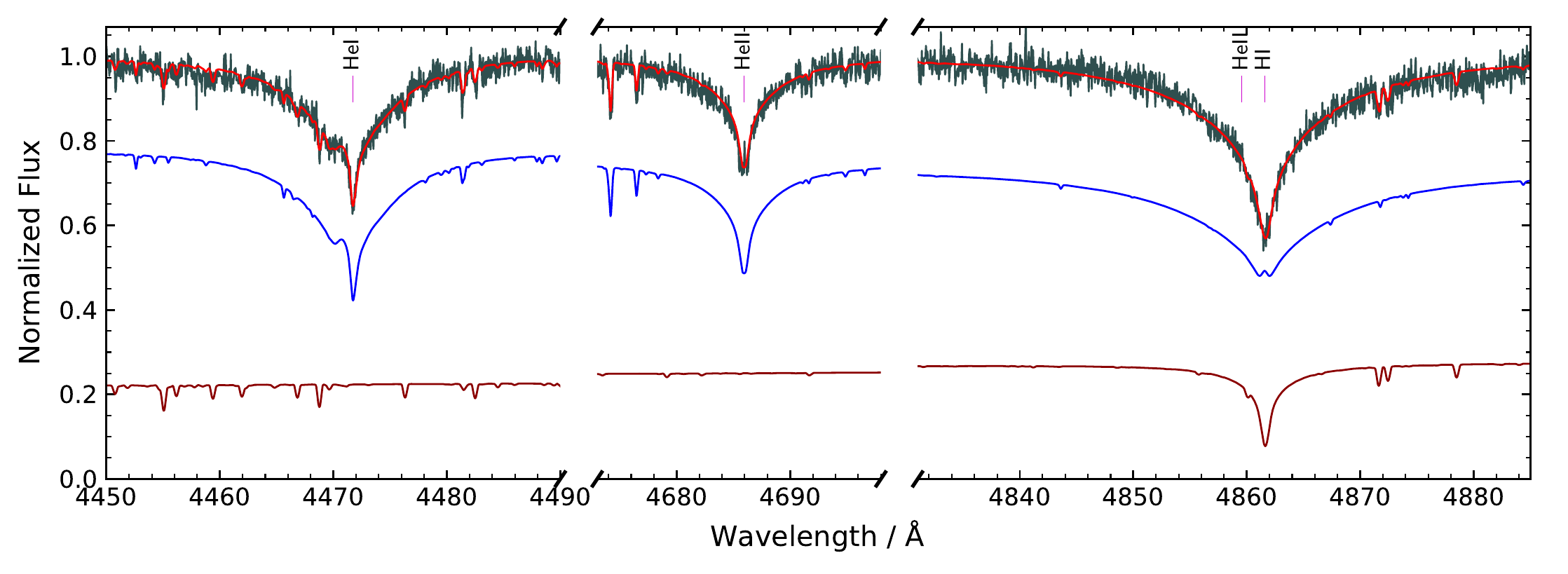}
\includegraphics[width=0.99\textwidth]{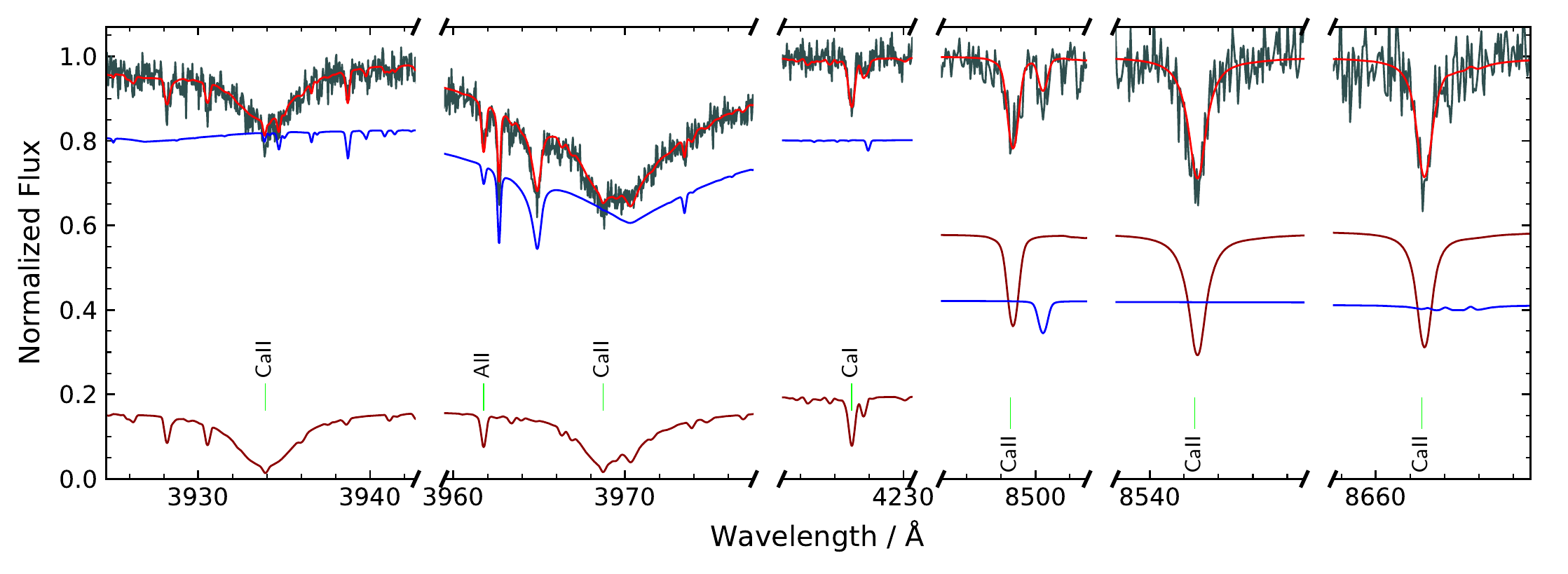}
\caption{Top: Examples of prominent helium and hydrogen lines in an individual UVES spectrum of \EC\ (grey). The combined model spectrum (red) is the sum of the contributions of the sdOB (blue) and F-type star (dark red). Bottom: Similarly, the strongest calcium lines in blue and red UVES spectra.}
    \label{fig:EC:atmlines}
\end{figure*}

To model the contribution of the sdOB component, we used fully line-blanketed, plane-parallel, homogeneous, hydrostatic, non-local thermodynamic equilibrium (NLTE) synthetic spectra computed with \tlusty\ and \synspec, in particular the most recent public versions \cite[for detailed descriptions, see][]{hubeny17a,hubeny17b,hubeny17c}.
Heavy metals were included in LTE in the spectrum synthesis as described by \cite{Latour2019b}. 
A first \up{estimate of} the atmospheric parameters was obtained using a large grid of synthetic spectra that was computed using the model atoms of \cite{Lanz2003}. 
This grid covers the full He-sdO/B parameter space in \teff, \logg, and helium abundance $\log n(\mathrm{He}) / n(\mathrm{H})$. 
A smaller grid was then constructed around the best-fit parameters for the sdOB obtained with the first grid.
To be consistent with the models used for the metal abundance analysis performed in Sect.~\ref{sect:metal}, this second grid uses the largest model atoms distributed with \tlusty\ 205.
Because these model atoms include more energy levels, optical transitions that involve high-lying levels were treated in NLTE.
The grid considers H, He, C, N, O, Ne, Mg, Al, Si, P, S, Ar, Ca, Fe, and Ni in NLTE at abundances close to those derived for the sdOB. 
The abundances of metals that could not be determined from the available spectra (most importantly Fe and Ni) were set to the values \uh{derived} for the iHe-sdOB HZ\,44 \citep{Dorsch2019}. 

Similarly, the cool companion was initially modelled using a large grid of model spectra.
Here, we used LTE model atmospheres and synthetic spectra computed with \A12\ \citep{Kurucz1996} and \synthe\ \citep{Kurucz1993}.
The initial grid extends from \teff\ = 4000 to 8000\,K, \logg\ = 2.0 to 5.2, \Z\ = $-2.0$ to $+0.5$\,dex, and covers microturbulent velocities of $v_\mathrm{tb}=0$, 1, and 2\,km\,s$^{-1}$.
A solar helium abundance was assumed.
The F-type companion is metal poor (\Z\,=\,$-1.9$) and strongly alpha enhanced, which is typical for halo stars \citep[e.g.][]{Fuhrmann1998}. %
The initial grid was therefore computed using a fixed alpha enhancement of $[\alpha/\mathrm{Fe}] = 0.4$\,dex relative to the photospheric solar values of \cite{asplund09}.
As for the sdOB component, a second, smaller grid was constructed around the best-fit parameters.
This grid also uses \A12  and \synthe\ model spectra, but it was additionally allowed to vary in alpha enhancement. 
\up{The dimensions of all four grids of synthetic spectra used in this analysis are summarised in Table \ref{tab:grids}.}

It is challenging to determine the surface gravity of the F-type companion accurately.
The wings of hydrogen lines, which are typically used to determine the surface gravity of sdB stars, are less useful for F-type stars because they are less sensitive to the density and correlate strongly with temperature.
In \EC, this is further complicated by the contribution of the sdOB, which has very broad hydrogen lines.
The strength of the bluest Balmer and Paschen lines is sensitive to the surface gravity due to level dissolution.
Unfortunately, our spectra lack coverage of these high Balmer lines, while the high Paschen lines are below the detection limit.
The observed Balmer series, from H$_\alpha$ up to H$_{11}$, excludes surface gravities $\log g_\mathrm{B}$\,<\,$4.0$.
Instead, the surface gravity of cool (F/G-type) stars is often derived from the \ion{Fe}{i-ii} ionisation equilibrium, which also depends on the effective temperature.
The strengths of the \ion{Mg}{i} triplet and the \ion{Ca}{ii} 3934, 3968\,\AA\ resonance lines are sensitive to the surface gravity and the respective abundances. %
As shown in Fig.~\ref{fig:EC:MgI}, the \ion{Mg}{i} triplet in \EC\ is well reproduced at $\log g_\mathrm{B}$\,=\,$4.7$, assuming an alpha enhancement of 0.4\,dex.

We performed global spectral fits in order to consider all sensitive absorption lines in the observed spectra. 
\up{Examples for strong hydrogen, helium, and calcium lines are shown in Fig.~\ref{fig:EC:atmlines}.}
The atmospheric parameters of both components and the surface ratio were varied simultaneously.
The individual UVES and HRS exposures were not stacked but evaluated at the same time.
This is necessary because the radial velocity difference between both components is not constant over more than a few days.
Spectral regions that were not well reproduced were removed before performing the final fit.
This includes metal lines with uncertain atomic data, as well as the cores of \ion{Ca}{ii} resonance and hydrogen Balmer lines, which are poorly modelled in our LTE models for the cool component.
After a first fit using large model grids, a second global fit was performed that used tailored grids for both components, as described above. 

The atmospheric parameters derived from the final spectral fit are listed in Table \ref{tab:EC:param}. 
They are consistent with the results obtained from the SED fit in Sect.~\ref{sect:sed}, but more precise.
We use 1-$\sigma$ intervals for the statistical uncertainties. 
Due to the high resolution and high total S/N of our spectra, purely statistical uncertainties are small. 
The total uncertainties are dominated by systematic effects, such as deficiencies in the synthetic spectra or limited accuracy in the normalisation of our spectra.
We estimate systematic uncertainties of 1\% in \teff, as well as 0.04 in $\log g_\mathrm{A}$, $\log n(\mathrm{He}) / n(\mathrm{H})$, \Z$_\mathrm{B}$, and \alp$_\mathrm{B}$.
These systematic uncertainties were added in quadrature to the smaller statistical errors.  
We used a higher systematic uncertainty of 0.10\,dex for the surface gravity of the cool component, $\log g_\mathrm{B}$, for two reasons.
First, there is no spectral feature that is strongly dependent on $\log g_\mathrm{B}$. 
Second, $\log g_\mathrm{B}$ is strongly correlated with other free parameters such as $T_\mathrm{eff,B}$, the surface ratio, and the alpha enhancement. 

We find a significant projected rotational velocity of $15.3\pm 0.2$\,\kms\ for the sdF-type companion. 
This rotation is relatively slow compared to the values \cite{Vos2018} found for the cool companions in their sample of nine long-period sdB + F/G/K-type systems.
\EC\,A is consistent with no rotation, which is not unusual since most hot \led{subdwarf}s are slow rotators, including stars found in wide binaries \citep{Geier2012}.

\section{Mass, radius, and luminosity}
\label{sect:mass}

\begin{table}
\centering
\caption{Stellar parameters for \EC\ as derived by combining the SED, spectroscopic, and parallax measurements. The mode and the highest density interval of each quantity are given for 1-$\sigma$ probability \citep[see][]{bailer18}.}
\label{tab:stellar}
\setstretch{1.25}
\begin{tabular}{lcc}
\toprule
\toprule
{} & A & B \\
\midrule
$R/R_\odot$ & $0.132\pm0.007$ & $0.75\pm0.07$ \\
$M/M_\odot$ & $0.40\pm0.06$  & $0.84^{+0.29}_{-0.23}$ \\
$L/L_\odot$ & $32\pm4$  & $0.74^{+0.15}_{-0.14}$ \\
\bottomrule
\end{tabular}
\end{table}

\begin{figure}%
\centering
\includegraphics[width=0.99\columnwidth]{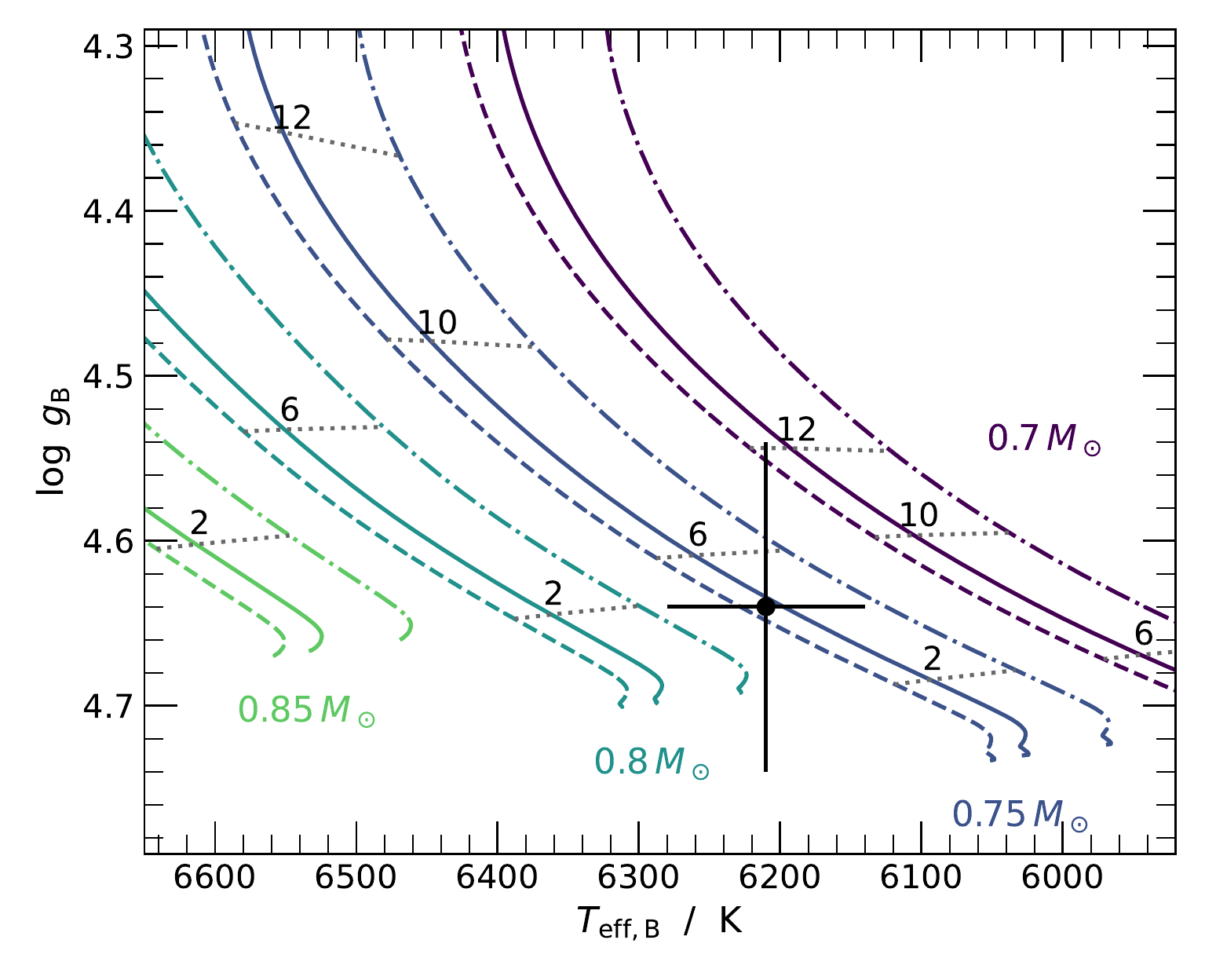}

\includegraphics[width=0.99\columnwidth]{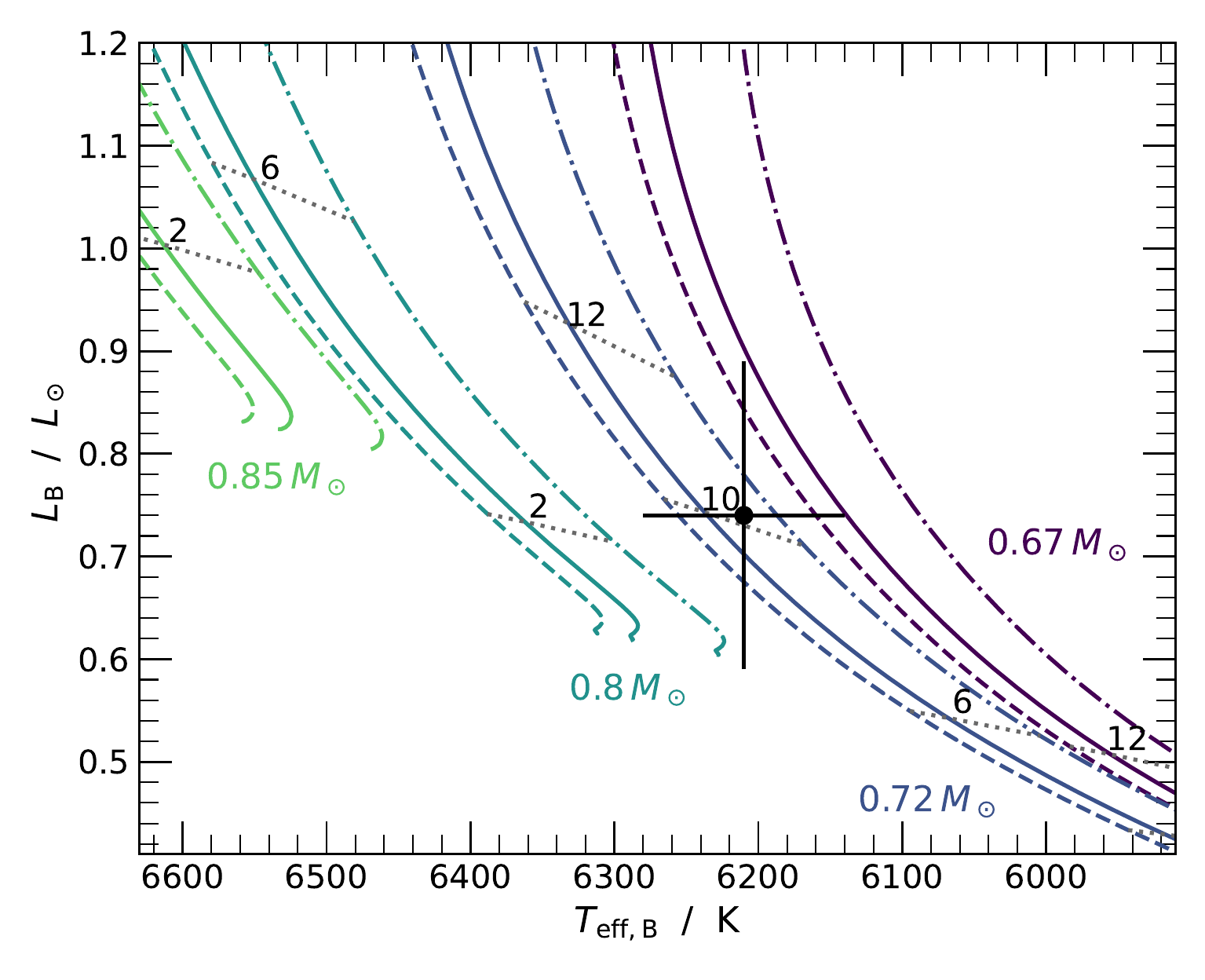}
\caption{Main-sequence evolutionary tracks from MIST \citep{Choi2016} for \Z\ = $-2.1$ (dashed), $-1.9$ (solid), and $-1.6$ (dash-dotted), as well as for four initial masses.
The upper panel shows the Kiel diagram, while the Hertzsprung-Russell diagram is shown in the lower panel. 
The dotted grey lines indicate equal age and are labelled in Gyr.
The parameters of \EC\,B and their uncertainties are marked by the cross.
}
\label{fig:MIST}
\end{figure}

Stellar parameters can be derived using the precise parallax measurement provided by the Early Data Release 3 (EDR3) of the \textit{Gaia} mission \citep{Gaia2016,Gaia2020}, $\varpi$\,=\,$1.40\pm0.04$\,mas.
We corrected the parallax for its zero-point offset following \cite{Lindegren2021} and inflated the corresponding uncertainty using the function suggested by \cite{El-Badry2021}. %
Stellar radii are then given as $R$\,=\,$\Theta/(2\varpi)$.
Here, $\Theta$ refers to the individual angular diameters determined from the SED.
In contrast to the SED fit performed in Sect.~\ref{sect:sed}, the angular diameter used here, $\log \Theta_\mathrm{A}/{\rm rad}$\,=\,$-11.078^{+0.018}_{-0.022}$, was obtained while keeping all atmospheric parameters fixed to the more precise spectroscopic values as listed in Table\,\ref{tab:EC:param}.
Combining the radii with the spectroscopic surface gravities then allowed us to derive the stellar masses $M$\,=\,$gR^2/G$, where $G$ is the gravitational constant.
Stellar luminosities are given as $L/L_\odot$\,=\,$(R/R_\odot)^2(T_\mathrm{eff}/T_{\mathrm{eff},\odot})^4$.
The stellar parameters for both components are listed in Table \ref{tab:stellar}.
We used a Monte Carlo method to propagate the uncertainties.

Due to the high uncertainty in surface gravity and surface ratio, the mass for the sdF component is poorly constrained. 
As discussed in Sect.~\ref{sect:velocity}, it is likely that the \EC\ system formed through \up{Roche-lobe} overflow.
However, the transferred mass is expected to be less than 0.03\,\msun\ \citep{Vos2018}.
It is likely that the transferred mass for \EC\,B is significantly below this value, given its relatively slow projected rotation. 
We can therefore derive an evolutionary mass for the sdF based on its spectroscopic effective temperature and surface gravity. 
Figure \ref{fig:MIST} shows single-star evolutionary tracks from the MIST project \citep{Choi2016} for metallicities of $\mathrm{[Fe/H]}$\,=\,$-1.9 \pm 0.3$. 
In the Kiel diagram \up{(Fig.~\ref{fig:MIST}; upper panel)}, our atmospheric parameters of \EC\,B are consistent with masses between about 0.8 and 0.7\,\msun. 
We cannot estimate an evolutionary age directly from the atmospheric parameters since our surface gravity puts \EC\,B close to the predicted zero-age MS. 
Given the low metallicity of \EC\,B, one would expect an age of the order of about 10\,Gyr or more. 
The spectroscopic surface gravity for the sdF may therefore be slightly overestimated. 

It is useful to consider the position of \EC\,B in the Hertzsprung-Russell  diagram (luminosity vs. \teff), given that the luminosity \uh{is (almost) independent of the spectroscopic surface gravity.}
As shown in the bottom panel of Fig.~\ref{fig:MIST}, our estimates for the luminosity and effective temperature of the sdF are consistent with a lower evolutionary mass, about 0.7\,\msun.
While this mass would be consistent with the expected age, the uncertainties are large.

It is not possible to directly derive an evolutionary mass for the sdOB because too many evolutionary tracks cross its position in the \teff\ - \logg\ plane. %
However, the mass expected for a sdOB that was formed through RLOF is close to the core mass that is required for the helium-flash at the top of the RGB, or about 0.49\,\msun\ at $\mathrm{[Fe/H]}\approx-2$ \citep{dor93}.
Although higher than the $0.40\pm 0.06$\,\msun\ found for \EC\,A, the lower value observed is still consistent with it being a core helium burning star.

\section{Metal abundance analysis}\label{sect:metal}
\begin{table}
\centering
\caption{Metal abundance results for \EC\,A by number fraction ($\log \epsilon = \log n_\mathrm{X} / \sum_i n_i$) and number fraction relative
to solar \cite[$\log \epsilon/\epsilon_\odot$,][]{asplund09}.
The number of resolved lines used per ionisation stage is given in the last column (with equivalent widths $>$\,10 m\AA).
}
\label{tab:EC:abu}
\vspace{-10pt}
\setstretch{1.1}
\begin{center}
\begin{tabular}{l@{\hspace{2pt}}rrr}
\toprule
\toprule
 Element  & \multicolumn{1}{c}{$\log \epsilon$} & \multicolumn{1}{c}{$\log \epsilon/\epsilon_{\odot}$} & \multicolumn{1}{c}{$N_\mathrm{lines}$}\\
\midrule
\ion{C}{ii-iv}         &  $-2.88\pm0.20$ &  $0.73\pm0.21$ & 19/45/2 \\
\ion{N}{ii-iii}        &  $-3.73\pm0.20$ &  $0.48\pm0.21$ & 22/9 \\
\ion{O}{ii-iii}        &  $-3.46\pm0.20$ &  $-0.11\pm0.21$ & 46/4 \\
\ion{Mg}{ii}           &  $-5.07\pm0.30$ &  $-0.64\pm0.30$ & 1 \\
\ion{Si}{iii-iv}       &  $-5.52\pm0.25$ &  $-0.99\pm0.25$ & 1/2 \\
\ion{P}{iii-iv}        &  $-5.92\pm0.30$ &  $0.70\pm0.30$ & 1/1 \\
\ion{S}{iii-iv}        &  $-5.76\pm0.30$ &  $-0.84\pm0.30$ & 1/1 \\
Ar                     &  <$-5.80^{+0.40}_{}$ &  <$-0.17^{+0.41}_{}$ \\
Ca                     &  <$-5.42^{+0.40}_{}$ &   <$0.27^{+0.40}_{}$ \\
Ti                     &  <$-5.84^{+0.50}_{}$ &   <$1.24^{+0.50}_{}$ \\
Fe                     &  <$-4.30^{+0.30}_{}$ &   <$0.23^{+0.30}_{}$ \\
Zn                     &  <$-5.36^{+0.40}_{}$ &   <$2.11^{+0.40}_ {}$ \\
Ga                     &  <$-5.82^{+0.40}_{}$ &   <$3.18^{+0.40}_{}$ \\
Ge                     &  <$-5.89^{+0.40}_{}$ &   <$2.50^{+0.40}_{}$ \\
Kr                     &  <$-4.89^{+0.40}_{}$ &   <$3.90^{+0.40}_{}$ \\
Sr                     &  <$-5.27^{+0.40}_{}$ &   <$3.90^{+0.40}_{}$ \\
Y                      &  <$-5.76^{+0.40}_{}$ &   <$4.07^{+0.40}_{}$ \\
Zr                     &  <$-6.71^{+0.40}_{}$ &   <$2.75^{+0.40}_{}$ \\
Sn                     &  <$-6.44^{+0.40}_{}$ &   <$3.56^{+0.40}_{}$ \\
\ion{Pb}{iii-iv}       &   $-4.01\pm0.30$ &    $6.27\pm0.32$ & 5/7 \\
\bottomrule
\end{tabular}
\end{center}
\end{table}

The atmospheric parameters of both components were kept fixed for the metal abundance analysis. 
The spectrum of the cool companion was modelled using \A12/\synthe\ using the final best-fit atmospheric parameters.
We used the global $\chi^2$ fitting procedure developed by \cite{Irrgang2014} to simultaneously determine the abundances for all metals that show sufficiently strong lines in the hot component (C, N, O, Mg, Si, P, Pb).
A grid of synthetic spectra that include lines of one metal only was computed for each metal using \synspec, always based \up{on the} same \tlusty\ atmosphere. 
This model atmosphere was calculated for the best-fit atmospheric parameters derived in Sect.~\ref{sect:atm}.  
\up{To allow an estimation of the microturbulent velocity, each grid was calculated for microturbulent velocities of 0 and 3 \kms.}
The full synthetic spectrum was then constructed by multiplication of all individual metal spectra, which were interpolated to the desired abundances.
This method is well tested for sdB and other B-type stars \citep[e.g.][]{Schaffenroth2021,Irrgang2020}.
It assumes that small changes in the metal abundances do not influence the atmospheric structure and that there are few intrinsic blends between lines of different metals.
The abundance fitting procedure was repeated using a \tlusty\ atmosphere that consistently includes the abundances from a first fit.
As before, spectral regions that were not well reproduced were removed from the fit.

\begin{figure*}%
\centering
\includegraphics[width=0.99\textwidth]{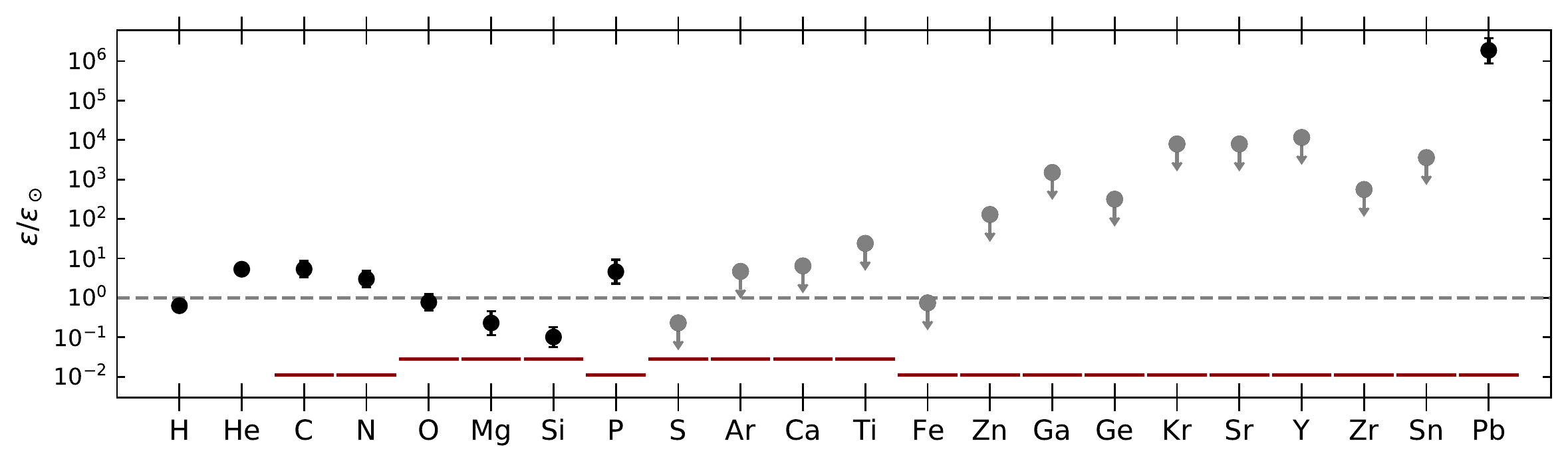}
\caption{Photospheric abundances for \EC\,A relative to solar values from \cite{asplund09}.
Abundance measurements are shown as black dots, while upper limits are marked with grey arrows.
Solid red lines show the corresponding metal abundances adopted in our model for \EC\,B\up{, as given by the best-fit \Z\ and \alp.}
The solar reference is indicated by the dashed grey line. 
}
\label{fig:EC:abu}
\end{figure*}

\up{Because the observed lines that originate from the sdOB component show no strong signs of a microturbulent velocity, the best-fit $v_\mathrm{tb,A}=2.1\pm 0.2$\,\kms\ can be considered as an upper limit. }
The final abundance pattern for \EC\,A is listed in Table \ref{tab:EC:abu} and shown in Fig.~\ref{fig:EC:abu}.
Upper limits were derived by eye.
They are stated as best-fit values, with an uncertainty that indicates at which abundance the predicted lines become clearly too strong.
The analysis of individual metal abundances is described in the following.

Plenty of strong \ion{C}{ii-iv}, \ion{N}{ii-iii}, and \ion{O}{ii-iii} lines are present in the spectrum of  \EC\,A.
The \ion{C}{ii} atom was updated to use resonance-averaged photo-ionisation cross-sections using data from TOPbase \citep{TOPbase}. %
This slightly changes the strengths of \ion{C}{ii} lines, but it has little effect on the general atmospheric structure because \ion{C}{ii} represents a small fraction of all carbon ions throughout the atmosphere (<\,2\,\%).
The sdOB star is enriched in carbon and nitrogen and has an approximately solar oxygen abundance.
The carbon, nitrogen, and oxygen lines in the spectrum of the F-type companion are too weak to be detected.
The \ion{Mg}{ii} 4481\,\AA\ doublet is present in both components, but the contribution of the sdOB is larger, despite its sub-solar magnesium abundance.
This is  because the flux contribution of the sdOB star is more than twice that of the F-type companion in this specific range.
The derived silicon abundance for the sdOB, about one-tenth solar, is mostly based on the strong \ion{Si}{iv}\,4088.9, 4116.1\,\AA\ lines and the weaker \ion{Si}{iii}\,4552.6\,\AA\ line.
The only detected silicon line that originates from the cool companion, \ion{Si}{i}\,3905.52\,\AA, is consistent with an alpha-enhancement of 0.4\,dex.
Two weak phosphorus lines in the spectrum of the sdOB, \ion{P}{iii} 4222.2\,\AA\ and \ion{P}{iv} 4249.7\,\AA, are clearly identified.
They are best reproduced at an abundance of about five times solar.
All detected calcium lines originate from the cool component.
Most notably, the \ion{Ca}{i} 4226.7\,\AA\ resonance line and the \ion{Ca}{ii} 8498, 8542, 8662\,\AA\ triplet are well reproduced at an alpha-enhancement of 0.4\,dex.

\begin{figure*}%
\centering
\includegraphics[width=0.99\textwidth]{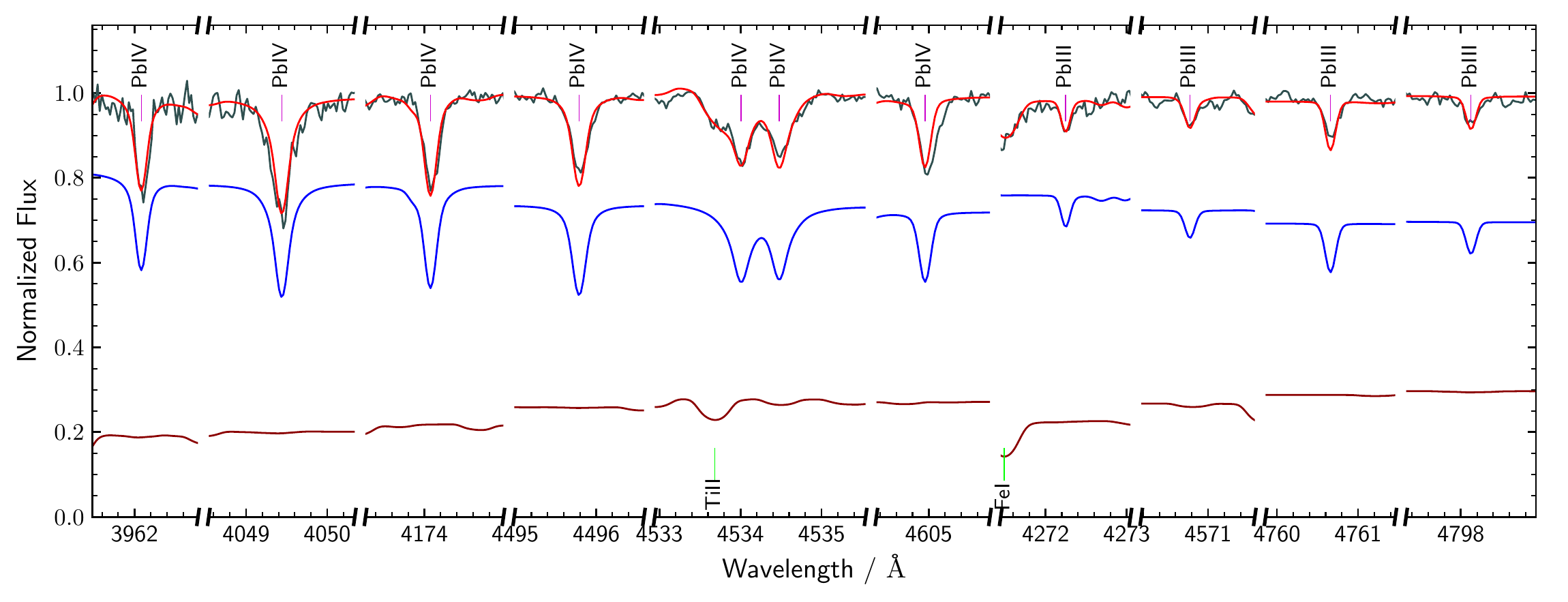}
\caption{Very strong lead lines in the HRS spectrum of \EC\ (grey).
The combined model spectrum (red) is the sum of the contributions of the sdOB (blue) and F-type star (dark red).}
    \label{fig:EC:Pb}
\end{figure*}

\begin{table}
\caption{
Lead lines detected in the spectrum of \EC.
References for the oscillator strengths are stated in the last column.
}
\label{tab:EC:Pb}
\vspace{-14pt}
\setstretch{1.08}
\begin{center}
\begin{tabular}{lccc}
\toprule
\toprule
Ion & $\lambda$ / \AA\ & $\log gf$ & Ref.  \\
\midrule
\ion{Pb}{iii} & 3854.080 & $+0.302$ & 1 \\
\ion{Pb}{iii} & 4272.660 & $-0.462$ & 1 \\
\ion{Pb}{iii} & 4571.219 & $+0.029$ & 1 \\
\ion{Pb}{iii} & 4761.120 & $+0.012$ & 1 \\
\ion{Pb}{iii} & 4798.590 & $-0.356$ & 1 \\
\ion{Pb}{iv} & 3962.467 & $-0.047$ & 2 \\
\ion{Pb}{iv} & 4049.832 & $-0.065$ & 2 \\
\ion{Pb}{iv} & 4174.478 & $-0.444$ & 3 \\
\ion{Pb}{iv} & 4496.223 & $-0.437$ & 3 \\
\ion{Pb}{iv} & 4534.447 & $+1.190$ & 3 \\
\ion{Pb}{iv} & 4534.917 & $+1.102$ & 3 \\
\ion{Pb}{iv} & 4605.400 & $-0.991$ & 3 \\
\bottomrule
\end{tabular}
\end{center}
\vspace{-10pt}
\tablefoot{
(1) \cite{alonso-medina09};
(2) \cite{safronova04}; 
(3) \cite{alonso-medina11}.
}
\end{table}

As shown in Fig.~\ref{fig:EC:Pb}, strong \ion{Pb}{iv} lines are present at rest wavelengths of 3962.5, 4049.8, 4174.5, 4496.2, 4534.4, 4534.9, and 4605.4\,\AA.
Although weaker, \ion{Pb}{iii} lines at 3854.1, 4272.7, 4571.2, 4761.1, and 4798.6\,\AA\ are clearly detected.
All identified lead lines are listed in Table \ref{tab:EC:Pb}. 
The 2019 HRS spectra were shifted to the rest frame of the sdOB and co-added to update the wavelengths of newly observed lead lines.
\up{This co-added spectrum was not used for the abundance fit, which, as before, was performed using all individual spectra.}
All modelled lead lines are reasonably well reproduced at an abundance between one and two million times solar, or 100 million times larger than the scaled solar value using the metallicity deduced from the cool companion.
This is significantly more than the 4.8\,dex enrichment derived by \cite{Jeffery2019}, who were unaware of any flux contribution from a companion.

Table \ref{tab:EC:abu} also lists upper limits for several elements\uh{, which had been detected in other iHe\,sdOBs, in particular \feige\ and \lsiv\ \citep{Dorsch2020}.}
Since no lines from these elements are detected, any 'best-fit' abundance obtained from a $\chi^2$ fit would strongly depend  on the location of the continuum.
We therefore prefer to obtain upper limits by eye.
All lines used to derive upper limits for \EC\,A up to iron are known to be well reproduced in our models of other He-sdOB stars. 
For sulphur, we used the weak \ion{S}{iv} 4485.7, 4504.2\,\AA\ lines, which seem to be just below the detection limit of the HRS and UVES spectra of \EC.
The only usable predicted argon line is \ion{Ar}{iii} 4183.0\,\AA.
The upper limits for calcium and titanium are based on \ion{Ca}{iii} 4233.7, 4240.7\,\AA,\ and \ion{Ti}{iv} 4618.2, 5398.9, 5492.5\,\AA, respectively.
The upper limit for iron is based on the non-detection of \ion{Fe}{iii} 4164.7, 4304.8, and 4310.4\,\AA.
In addition to lead, we also searched for heavy metals that have been detected in optical spectra of the intermediate He-sdOBs \feige\ and \lsiv: Zn, Ga, Ge, Kr, Sr, Y, Zr, and Sn.
Upper limits for these elements are based on the non-detection of
\ion{Zn}{iii} 4818.9, 5075.2\,\AA,
\ion{Ga}{iii} 4380.6, 4381.8, 4993.9\,\AA, \ion{Ge}{iii} 4179.1, 4260.9\,\AA, \ion{Kr}{iii} 4067.4, 4226.6\,\AA, \ion{Sr}{iii} 3936.4\,\AA, \ion{Y}{iii} 4039.6, 4040.1\,\AA,  \ion{Zr}{iv} 4198.3, 5462.4\,\AA,  and \ion{Sn}{iv} 4216.2\,\AA.
We used the same atomic data as \cite{Dorsch2020} for these ions.
The resulting upper limits rule out extreme enrichments as observed for lead, but would still be consistent with strong enrichment compared to solar or even mean sdB values.
Ultraviolet spectra would enable us to determine abundances for most of these heavy metals, but they are not presently available.
A small number of lines in the UVES and HRS spectra remain unidentified.
The identification of these lines is complicated by the composite nature of \EC, since the lines of the F-type companion are only slightly more broadened by rotation than those of the sdOB.
Table \ref{tab:EC:unid} lists the rest wavelengths of all detected unidentified lines, assuming that they originate from the sdOB component.
We only list lines that are present in both coadded HRS and coadded UVES spectra, or are strong enough to be identified in single exposures.

The overall abundance pattern for \EC\,A is similar to that derived by \cite{Jeffery2019}, but shifted to higher abundances due to the contribution of the sdF star (unaccounted for previously).
Carbon, nitrogen, and phosphorus are enhanced \led{relative} to solar values, while the oxygen abundance is about solar.
Magnesium, silicon, and sulphur are sub-solar.
With respect to the primordial metallicity plus alpha enhancement, all detected elements would be enhanced.
Similar patterns for these light metals have been observed in other heavy-metal iHe-sdOBs: the zirconium-rich \feige\ and \lsiv\ \citep{naslim11,Dorsch2020}, the zirconium- and lead-rich HE\,2359-2844 and HE\,1256-2738 \citep{naslim13}, the lead-rich PG\,1559+048 and FBS\,1749+373 \citep{Naslim2020}, as well as PG\,0909+276 and UVO\,0512-08 \citep{Edelmann2003,Wild2018}, which are extremely enriched in iron-group elements.
Unlike the hotter lead-rich iHe-sdOBs UVO\,0825+15 \citep{Jeffery2017}, HZ\,44, and HD\,127493 \citep{Dorsch2019}, \EC\,A does not show the distinct CNO-cycle pattern (strong N, weak C and O). 
Notably, the metal abundance pattern is not shifted to lower values when compared to other heavy-metal iHe-sdOBs. 
This indicates that the metal abundance patterns observed for iHe-sdOB stars are not strongly dependent on the initial metallicity. 
The abundances derived for C, N, O, and Si are almost identical to those of HE\,1256-2738, an apparently single lead-rich iHe-sdOB found to be a Galactic halo member by \cite{Martin2017}.

\EC\,A does not seem to share the extreme zirconium abundance observed in some iHe-sdOBs or the extreme enrichment in iron-group elements with respect to the Sun found in others, although our upper limits are high. 
The latter may be due to the very low primordial metallicity of the system.
The lead enhancement of \EC\,A is the most extreme found in any hot \led{subdwarf}, or, to our knowledge, in any star.
The strong enrichment of heavy metals in the photospheres of heavy-metal sdOBs is usually discussed in terms of selective radiative levitation.
In this picture, the strong heavy metal lines observed in the emergent spectrum are the result of a chemically stratified envelope, in which a thin metal-rich layer overlaps with the line-forming region.
As mentioned by \cite{Jeffery2019}, atmospheric models that include a physical treatment of stratification by diffusion are required to estimate the total amount of lead in the enriched layers.
These models would then have to be compared with lines that form at various optical depths, ideally using both far-UV and optical spectra. 

The flux contribution of the sdF only overtakes that of the sdOB at about 7200\,\AA\ in our best-fit model. 
Therefore, and due to the low metallicity, relatively few metal lines that originate from the sdF component are detectable in our spectra.
Most of them are well reproduced at the best-fit metallicity and alpha-enhancement. 
The strongest metal lines detected include transitions in the \ion{Na}{i}, \ion{Mg}{i}, \ion{Al}{i}, \ion{Si}{i}, \ion{Ca}{i-ii}, \ion{Ti}{ii}, \ion{Cr}{i}, \ion{Mn}{i}, \ion{Fe}{i-ii}, and \ion{Ni}{i} atoms. 
The strontium lines \ion{Sr}{ii} 4077.7, 4215.5\,\AA\ seem to be well reproduced at the scaled solar abundance. 
We also detect \ion{Ba}{ii} 4554.0, 4934.1\,\AA, which are somewhat too weak in our models at the scaled solar abundance. 
This discrepancy may be due to a weak enrichment in barium, but may also be caused by deficiencies in our synthetic spectra, such as uncertain atomic data or NLTE effects. 
The existence of dwarf barium stars is usually explained with  pollution through wind accretion or RLOF from an AGB star  \citep{Jorissen1992, Gray2011}.
Given that \EC\,A is likely still on the \up{horizontal branch}, it seems unlikely that the sdF component is a barium star. 

It is likely that the present \EC\ system formed through mass transfer from a RGB star to the sdF companion.
One might therefore expect at least some pollution of the sdF companion by material processed in the CNO-cycle\todo{, although diluted by convection}. 
Unfortunately, no carbon, nitrogen, or oxygen lines that originate from the sdF are detectable in our spectra. 

\section{\uh{Analysis of the radial velocity curve}}
\label{sect:velocity}

\begin{figure}%
\centering
\includegraphics[width=0.49\textwidth]{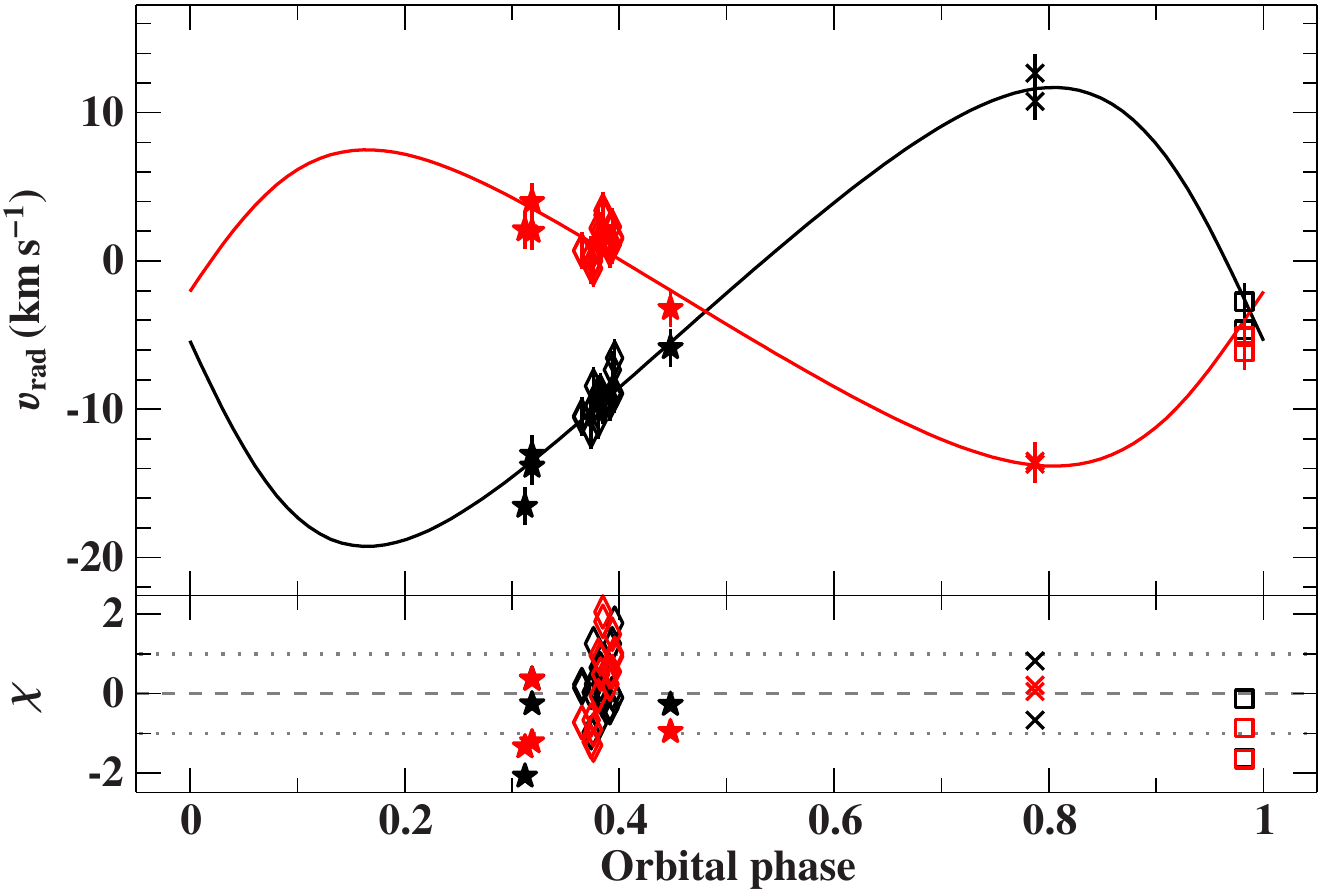}
\caption{Elliptic orbits fit to the radial velocities of both \EC\ A (black) and B (red). 
\uh{Asterisks} indicate UVES measurements, while HRS data from 2017, 2018, and 2019 are marked by crosses, squares, and diamonds, respectively.
}
\label{fig:EC:vrad}
\end{figure}

\begin{table}
\centering
\caption{Orbital parameters for \EC. \uh{The gravitational redshift $\varv_\mathrm{grav}$ is calculated from the stellar parameters listed in Table \ref{tab:stellar}}.}
\label{tab:EC:rvfit}
\renewcommand{\arraystretch}{1.25}
\begin{tabular}{lr}
\toprule
\toprule
Parameter & Value \\
\midrule
Period $P$ & $457.0^{+1.2}_{-1.5}$\,d \\
Epoch of periastron $T_{\mathrm{periastron}}$ & $56160^{+21}_{-17}$\,MJD \\
Eccentricity $e$ & $0.22^{+0.13}_{-0.08}$ \\
Longitude of periastron $\omega$ & $276^{+21}_{-18}$\,deg \\
Velocity semiamplitude $K_A$ & $15.5^{+1.7}_{-1.6}$\,km\,s${}^{-1}$ \\
Velocity semiamplitude $K_B$ & $10.7\pm1.4$\,km\,s${}^{-1}$ \\
Gravitational redshift $\varv_\mathrm{grav_A}$  & $1.95^{+0.22}_{-0.20}$\,km\,s${}^{-1}$  \\
Gravitational redshift $\varv_\mathrm{grav_B}$ & $0.72^{+0.21}_{-0.17}$\,km\,s${}^{-1}$  \\
Systemic velocity $\gamma$ & $-3.3\pm0.4$\,km\,s${}^{-1}$ \\
\midrule
Derived parameter & Value \\
\midrule
Mass ratio $q=K_B / K_A=M_A/M_B$ & $0.69^{+0.06}_{-0.05}$ \\ %
Projected semimajor axis $a_A \sin(i)$ & $0.63\pm0.05$\,au \\ %
Projected semimajor axis $a_B \sin(i)$ & $0.44\pm0.04$\,au \\ %
\bottomrule
\end{tabular}
\end{table}

A total of 27 HRS and UVES spectra are of sufficient quality to measure the radial velocities for both components, which are listed in Table \ref{tab:vrad}. 
These spectra cover a time span of \up{$\Delta t = 2880$ days}, but they were taken in \up{only five observing runs, the longest of which was just two weeks}.
This coverage is too irregular to precisely determine the orbital parameters of the \EC\ system. 
\up{We initially searched for orbital periods by fitting circular orbits to all available radial velocities, corrected for the gravitational redshifts. 
A unique best orbital period, $P$ $\approx$ 457 days, was obtained by finely sampling orbital frequencies between 1/330 and 1/550 d$^{-1}$ with steps of $0.01/\Delta t = 3.6\times10^{-6}$ $\mathrm{d}^{-1}$.
Fitting eccentric orbits resulted} in a somewhat eccentric orbit with \up{the same 457-day period} and velocity semi-amplitudes of $K_A=15.5\pm1.7$ km\,s$^{-1}$ and $K_B=10.7\pm1.4$ km\,s$^{-1}$ (see Table \ref{tab:EC:rvfit}). 
The phased radial velocity curves are shown in Fig.~\ref{fig:EC:vrad}.
More observations are required to \uh{improve the} orbital solution, in particular for the eccentricity.

Since the orbital period of \EC\ is certainly longer than 50 days, it is likely that the system was formed through stable \up{Roche-lobe} overflow. 
\cite{Vos2017} discovered a positive correlation between the eccentricity and orbital period for post-RLOF systems. 
Given the orbital period of about 457 days, one would expect a low eccentricity for the orbit of \EC. 
We find an eccentricity of $e=0.22^{+0.13}_{-0.08}$, which, however, strongly depends on the radial velocities derived from the HRS spectra taken in 2017. %

A correlation between orbital period and mass ratio of post-RLOF sdB+MS binaries was found by \cite{Vos2019}.
Subsequently, \cite{Vos2020} showed that the observed relation can be explained in terms of the system metallicity.  
Orbital periods of sdB+MS systems decrease with metallicity, because low-metallicity donor stars have smaller radii at the top of the RGB.
In order to produce a sdB, the mass transfer must happen close to the top of the RGB, so that the RGB star can ignite helium burning in its core.
Low-metallicity systems therefore have shorter initial periods, which leads to shorter final periods \uh{once the mass transfer has stopped}.
For halo systems at \Z\,=\,$-1.8\pm0.5$, \cite{Vos2020} predicted orbital periods of 300 to 500 days and mass ratios $q=M_\mathrm{sdB} / M_\mathrm{MS}$ of 0.6 to 0.8, which is consistent with our (preliminary) result \uh{for \EC, $q=K_B/K_A=M_A/M_B=0.69\pm 0.06$}. 
Assuming a canonical mass of $M_\mathrm{sdB}=0.49$\,\msun\ for the sdOB, \uh{the mass} ratio would put the mass of the sdF at $0.71\pm 0.06$\,\msun.

\section{Kinematics}
\label{sect:kinematics}

\begin{figure}%
\centering
\includegraphics[width=0.49\textwidth]{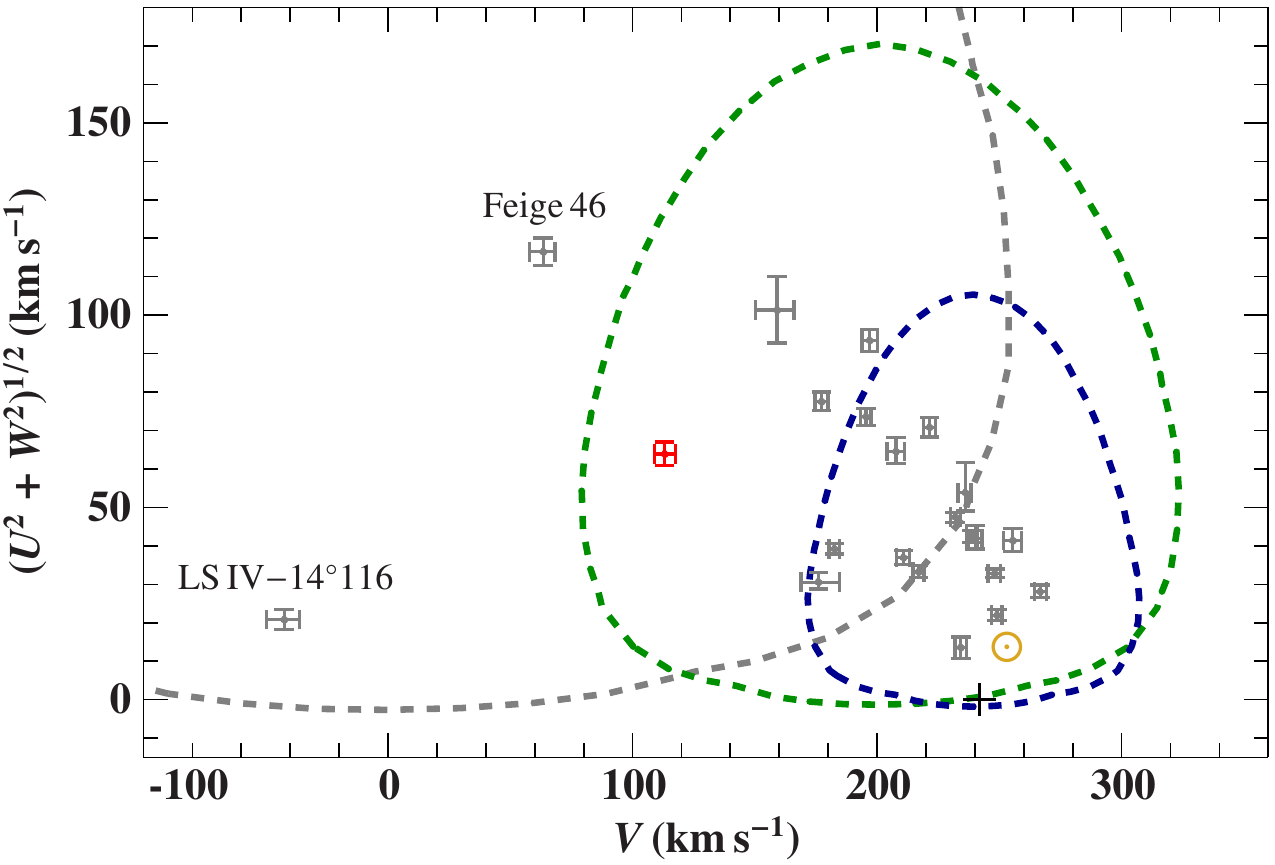}
\caption{
Toomre diagram showing space velocities with respect to the Galactic centre. \uh{The  velocity  component $V$ is  measured  in  the direction of the rotation of the Galaxy, $U$ towards the Galactic centre, and $W$ perpendicular to the plane.} 
The position of \EC, the Sun, and the local standard of rest (LSR) are marked by the red cross, yellow circled dot, and black plus sign, respectively.
The grey, green, and blue dashed lines indicate 2-$\sigma$ velocity dispersions from \cite{Robin2003} for the halo, thick disc, and old thin disc, respectively.
Stars from the sample of intermediate He-sdO/Bs studied by \cite{Martin2017}, updated using \textit{Gaia} EDR3 parallaxes and proper motions are shown in grey. %
Probable halo stars are labelled. 
}
\label{fig:EC:Toomre}
\end{figure}

Proper motions and the parallax from \textit{Gaia} EDR3 combined with the system's radial velocity can be used to derive the present 3D space velocity of \EC.
The Galactic orbit of \EC\ can then be traced back using a model for the Galactic potential: \todo{here, for 10 Gyr}. 
We used Model I of \cite{Irrgang2013} for the Galactic potential, which is a revision of the \cite{Allen1991} potential. 
Given the sparse coverage of our radial velocities, we \uh {adopted an} uncertainty of 5\,km\,s$^{-1}$ on the system's radial velocity. 
The resulting orbit for \EC\ has a low angular momentum perpendicular to the Galactic disc $J_Z = 930 \pm 40$ kpc\,km\,s$^{-1}$, but a relatively high eccentricity of $e = 0.53 \pm 0.02$.
\uh{}
The current velocity towards the Galactic centre $U$, perpendicular to the disc $W$, and in the direction of Galactic rotation $V$ can be used to place \EC\ in the Toomre diagram. \uh{When compared to the Toomre parameters predicted from Besan\c{c}on Galactic models \citep[][]{Robin2003}, the} location of \EC\ in this diagram is consistent with either a thick-disc or halo origin \uh{(see  Fig.~\ref{fig:EC:Toomre})}.
\todo{
Although most hot \led{subdwarf} stars seem to be part of the thin disc, several intermediate He-sdO/Bs in the sample of \cite{Martin2017} have been classified as thick discs or halos. 
To facilitate the comparison with \EC, we repeated the orbit calculation for iHe-sdOBs from this sample using reliable \textit{Gaia} EDR3 proper motions and parallaxes ($\sigma_\varpi/\varpi$\,<\,$10\,\%$).
The heavy-metal iHe-sdOBs \feige\ and HZ\,44 \uh{\citep{Dorsch2019, Dorsch2020}} were considered in addition. 
We corrected the proper motions provided by \textit{Gaia} EDR3 for bright stars \uh{($G<13$\,mag), following \citet {Cantat-Gaudin2021}}.
As before, we applied corrections to the parallax measurements \citep{Lindegren2021} and their uncertainty \citep{El-Badry2021}. 
The resulting orbital parameters are summarised in Table \ref{tab:kin}.
}

Given the low metallicity of \EC, the system may be part of the low-metallicity tail of the thick disc, also termed metal-weak thick disc (MWTD). 
According to \cite{Chiba2000}, the low-metallicity cut-off of the MWTD is close to \Z\,=\,$-2$, comparable to the metallicity of \EC\,B. 
At this low metallicity, the halo population in their sample dominates even at small distances from the Galactic disc, such as the $Z=-593^{+13}_{-16}$\,pc observed for \EC. 
However, is not possible to discern between a MWTD or halo origin for any particular system that shows disc-like kinematics \citep[see e.g.][]{Reddy2008}.

\section{Conclusions}
\label{sect:conclusions}

We performed a detailed analysis of high-resolution spectra of \EC and can confirm that the system contains an extremely lead-rich intermediate He-sdOB, as found by \cite{Jeffery2019}.
However, we find a second component in its spectrum: a strongly alpha-enhanced and metal-poor F-type \led{subdwarf}. 
\EC\,A is the first heavy-metal sdOB found in a binary system. 
Our updated metal abundances for the sdOB component are similar to those derived in the previous analysis, but shifted to higher abundances.
\EC\,A therefore remains the most lead-rich hot \led{subdwarf} known to date. 
Although the initial metallicity of the system is low, the abundances for the hot component are \uh{quite similar to those of} other iHe-sdOB stars, \uh{some of which probably have significantly higher initial metallicities, given that they belong to the younger thin-disc population (such as PG\,1559+048, see Sect. \ref{fig:EC:Toomre})}. 
The observed abundance pattern of \EC\,A is likely the result of strong diffusion processes.
Ultraviolet spectroscopy would enable us to derive a more complete abundance pattern for the sdOB component. 

\EC\,A is the first hot \led{subdwarf} found in a long-period binary with a known metallicity \Z\,<\,$-1$. 
The low metallicity (\Z$_\mathrm{B}$\,=\,$-1.95$) and strong alpha enhancement (\alp$_\mathrm{B}$\,=\,0.4) derived for the sdF component indicate that the system is part of the Galactic halo or metal-weak thick disc.
The system is therefore likely old \citep[$\gtrsim$10\,Gyr, e~g.][]{Helmi2020}. 
Other hot \led{subdwarf}s with a low known metallicity are part of globular clusters, but they are typically not found in binary systems \citep{Latour2018}. 
Still, the stellar evolution models of \cite{Han2008} predict a significant fraction of old sdO/Bs to have been formed through stable RLOF, which seems to be the case for \EC.

To determine the stellar parameters mass, radius, and luminosity for both components, we combined the parallax provided by \textit{Gaia} EDR3 with the angular diameter derived from a SED fit and our spectroscopic atmospheric parameters.  
The resulting mass for \EC\,A, $0.40\pm 0.06$\,\msun, is consistent with the range of masses expected for hot \led{subdwarf} stars on the \up{extreme horizontal branch}.
The spectroscopic mass for \EC\,B, $0.84^{+0.29}_{-0.23}$\,\msun, is associated with a high uncertainty because its surface gravity is not easily determined.
In fact, the comparison of MIST evolutionary tracks with our effective temperature and luminosity points to a lower mass for the sdF, about 0.7\,\msun.
This mass would be consistent with an evolutionary age of about 10\,Gyr.

Radial velocity variations suggest that the system is likely a wide binary with an orbital period of about 457 days. 
An eccentric fit to the radial velocities results in a mass ratio $q=M_A/M_B=0.69\pm 0.06$.
\cite{Vos2019} recently found a strong \uh{relation} between the orbital period and mass ratio for long-period hot \led{subdwarf} binaries.
This relation could be explained by a correlation between the final orbital period and metallicity of such systems \uh{found} by \cite{Vos2020}, which results from different radii \uh{of the progenitors} at the \uh{tip} of the RGB\uh{, when mass transfer started.} 
Given the low metallicity of \EC, the system helps to constrain these relations at much lower metallicity than before. 
The current results for the orbital period, mass ratio, and metallicity are consistent with the predictions of \cite{Vos2020} for post-RLOF systems.
Additional spectra that sample the full orbital period are required to obtain reliable orbital parameters. %

We also performed \uh{a kinematic} analysis, based on the system radial velocity determined from the current radial velocity curve.
\EC\ \uh{is} on an eccentric orbit around the Galactic centre, which is consistent with a halo or thick-disc membership.
Several heavy-metal iHe-sdOB stars have been found in the Galactic halo. 
From the \textit{Gaia} EDR3 data, we can confirm that the zirconium-rich \feige\ and \lsiv\ \citep{Latour2019a,Dorsch2020} are halo members.

\begin{acknowledgements}
We thank the referee for the careful reading and comments that improved this paper. 
A.I., U.H.\ \uh{and M.D.\ } acknowledge funding by the Deutsche For\-schungs\-gemeinschaft (DFG) through grants IR190/1-1, \uh{HE1356/70-1} and HE1356/71-1.
This research made use of the lmfit python package developed by \cite{lmfit}.
Based on observations obtained with the Southern African Large Telescope (SALT) under programmes 2017-1-SCI-004, 2018-2-SCI-033, and 2019-1-MLT-003. 
Based on observations collected at the European Southern Observatory under ESO programme 088.D-0364(A).
This work has made use of data from the European Space Agency (ESA) mission {\it Gaia} (\url{https://www.cosmos.esa.int/gaia}), processed by the {\it Gaia} Data Processing and Analysis Consortium (DPAC, \url{https://www.cosmos.esa.int/web/gaia/dpac/consortium}). Funding for the DPAC has been provided by national institutions, in particular the institutions participating in the {\it Gaia} Multilateral Agreement.
This publication makes use of data products from the Wide-field Infrared Survey Explorer, which is a joint project of the University of California, Los Angeles, and the Jet Propulsion Laboratory/California Institute of Technology, funded by the National Aeronautics and Space Administration. 
This research has made use of NASA's Astrophysics Data System.
The Armagh Observatory and Planetarium are funded by direct grant from the Northern Ireland Department for Communities.
\end{acknowledgements}

\bibliographystyle{aa}
\bibliography{EC.bib}

\begin{appendix}

\section{Additional material}
\label{appendix}

\begin{table}
\setstretch{1.2}
\captionof{table}{Dimensions for the four grids of synthetic spectra used for the spectroscopic and SED analyses.
For each parameter, the maximum and minimum values, as well as the step width are stated.
As described in Sect.~\ref{sect:atm}, the large and small \tlusty/\synspec\ grids were each computed using a fixed metal abundance pattern.}
\label{tab:grids}
\vspace{-11pt}
\begin{center}
\begin{tabular}{r r r r r}
\toprule
  &\multicolumn{2}{c}{\tlusty/\synspec} & \multicolumn{2}{c}{\A12/\synthe} \\
 \multicolumn{1}{c}{Parameter} &large & small & large & small \\ 
\midrule
\ldelim\{{3}{*}[\teff\ (K) ${}$]  &27500 & 37300 & 4000 & 6150 \\
 & 47500 & 38800 & 8000 & 6275 \\
 & 1250 & 500 & 200 & 25 \\[4pt]
\ldelim\{{3}{*}[\logg\ ${}$]&4.750 & 5.60 & 2.00 & 4.50 \\
&6.125 & 6.00 & 5.20 & 4.80 \\
&0.125 & 0.20 & 0.20 & 0.10 \\[4pt]
\ldelim\{{3}{*}[\logy\ ${}$]&$-1.00$ & $-0.23$ & $-1.07$ & $-1.07$ \\
&$+2.00$ & $+0.07$ & $-1.07$ & $-1.07$ \\
&0.25 & 0.15 & -- & -- \\[4pt]
\ldelim\{{3}{*}[\Z\ ${}$]& -- & -- & $-2.00$ & $-2.05$ \\ 
& -- & -- & $+0.50$ & $-1.85$ \\
& -- & -- & 0.50 & 0.20 \\[4pt]
\ldelim\{{3}{*}[\alp\ ${}$]& -- & -- & 0.40 & 0.33 \\
& -- & -- & 0.40 & 0.44 \\
& -- & -- & -- & 0.11 \\[4pt]
\ldelim\{{3}{*}[$v_\mathrm{tb}$ (\kms) ${}$]& 5.0 & 0.0 & 0.0 & 1.5 \\ 
& 5.0 & 3.0 & 2.0 & 2.5 \\ 
&-- & 3.0 & 1.0 & 1.0 \\
\bottomrule 
\end{tabular}
\end{center}
\end{table}

\begin{table}
\caption{
Unidentified lines in the HRS and UVES spectra of \EC. 
Equivalent widths are stated for the composite spectrum. 
Rest wavelengths assume that lines originate from the sdOB. 
}
\label{tab:EC:unid}
\vspace{-14pt}
\setstretch{1.08}
\begin{center}
\begin{tabular}{ccc}
\toprule
\toprule
$\lambda$ / \AA\ & EW / m\AA\ & Comment  \\
\midrule
4081.692 & 19 & sharp  \\
4182.414 & 11 & sharp  \\
4273.738 & 12 &  \\
4400.840 & 10 &  \\
4450.986 & 8 &  \\
4581.979 & 25 & broad  \\
4664.656 & 13 & broad \\
4802.251 & 12 & \\
5021.613 & 27 & artifact?\\
5094.107 & 15 & \\
5438.381 & 22 & \\
7298.346 & 78 & \\
\bottomrule
\end{tabular}
\end{center}
\end{table}

\begin{table}
\caption{
Radial velocities with barycentric correction applied.
The gravitational redshifts have not been corrected.
The typical uncertainty is of the order of 2\,\kms, depending on S/N.
There may be a small systematic trend in the radial velocities derived from HRS spectra that were taken in 2019.
}
\label{tab:vrad}
\vspace{-14pt}
\setstretch{1.08}
\begin{center}
\begin{tabular}{crrc}
\toprule
\toprule
MJD & $v_\mathrm{rad, B}$ / \kms & $v_\mathrm{rad, A}$ / \kms & Spectrograph  \\
\midrule
55846.1 &   4.0 & $-$14.6  &   UVES \\
55846.1 &   4.1 & $-$15.7  &   UVES \\
55849.1 &   6.0 & $-$12.2  &   UVES \\
55849.1 &   4.0 & $-$13.0  &   UVES \\
55908.0 &  $-$1.2 &  $-$5.0  &   UVES \\
57891.0 & $-$11.7 &  13.5  &   HRS  \\
57891.0 & $-$11.5 &  11.6  &   HRS  \\
58437.0 &  $-$3.3 &  $-$3.2  &   HRS  \\
58437.0 &  $-$4.2 &  $-$0.3  &   HRS  \\
58612.0 &   2.4 & $-$10.5  &   HRS  \\
58612.0 &   2.6 &  $-$9.4  &   HRS  \\
58616.0 &   1.9 &  $-$10.7  &   HRS  \\
58616.0 &   2.1 &  $-$11.0  &   HRS  \\
58617.0 &   2.0 &  $-$8.3  &   HRS  \\
58617.0 &   2.3 &  $-$8.8  &   HRS  \\
58619.0 &   4.1 &  $-$10.0  &   HRS  \\
58619.0 &   2.9 &  $-$7.5  &   HRS  \\
58620.0 &   2.5 &  $-$7.6  &   HRS  \\
58620.0 &   4.9 &  $-$8.0  &   HRS  \\
58621.0 &   3.9 &  $-$8.4  &   HRS  \\
58621.0 &   3.8 &  $-$9.0  &   HRS  \\
58624.0 &   3.4 &  $-$8.4  &   HRS  \\
58624.0 &   3.0 &  $-$7.4  &   HRS  \\
58625.0 &   3.4 &  $-$7.4  &   HRS  \\
58625.0 &   4.1 &  $-$7.3  &   HRS  \\
58626.0 &   3.3 &  $-$7.9  &   HRS  \\
58626.0 &   4.1 &  $-$7.2  &   HRS  \\
\bottomrule
\end{tabular}
\end{center}
\end{table}

\begin{sidewaystable*}
\caption{
\todo{Orbital parameters for a sample of intermediate He sdOB stars studied by \citep{Martin2017} and recent additions (HZ\,44, Feige\,46, \EC).
The quantities $R_\mathrm{max}$, $R_\mathrm{min}$, and $Z_\mathrm{max}$ refer to the maximum and minimum distance from the Galactic centre, as well as the maximum distance from the Galactic disc.
Known zirconium- or lead-rich stars are marked by $^\dagger$. }
}
\label{tab:kin}
\vspace{-14pt}
\setstretch{1.08}
\begin{center}
\begin{tabular}{lrrrrrrrrrrrrrrrrrr}
\toprule
\toprule
                     Star &    $v_\mathrm{rad}$ &  $\pm$ &   $R_\mathrm{max}$ &  $\pm$ &  $R_\mathrm{min}$ &  $\pm$ &      $Z_\mathrm{max}$ &  $\pm$ &     $V$ &  $\pm$ &      $U$ &  $\pm$ &     $W$ &  $\pm$ &    $J_Z$ &  $\pm$ &    $e$ &  $\pm$ \\
                      &    \kms\ &  &  kpc &   &  kpc &   &      kpc &   &     \kms\ &   &  \kms\ &   &     \kms\ &   &    Mpc\,km\,s$^{-1}$ &   &    &   \\
\midrule
   PG\,0909+276             &   $20.0$ &      2.0 &  9.23 &    0.18 & 8.59 &    0.05 & 0.42 &     0.03 & $249.1$ &  $2.1$ &  $-0.3$ &  $1.9$ &  $22.0$ &  $1.5$ &  $2.19$ &  $0.02$ & 0.04 & $0.01$ \\
     HD\,127493$^\dagger$   &  $-17.0$ &      3.0 &  8.42 &    0.08 & 7.60 &    0.13 & 0.10 &     0.01 & $234.2$ &  $2.4$ &  $13.3$ &  $2.5$ &  $-3.2$ &  $1.8$ &  $1.98$ &  $0.02$ & 0.05 & $0.01$ \\
   UVO\,0512$-$08           &   $11.0$ &      3.3 &  9.46 &    0.20 & 8.24 &    0.07 & 0.44 &     0.03 & $248.1$ &  $2.5$ & $-21.0$ &  $2.9$ & $-25.2$ &  $1.6$ &  $2.17$ &  $0.03$ & 0.07 & $0.02$ \\
  HE\,1238$-$1745           &   $-7.9$ &      2.8 &  9.18 &    0.15 & 7.19 &    0.12 & 0.96 &     0.07 & $240.2$ &  $2.8$ &  $39.1$ &  $3.1$ &  $15.6$ &  $2.1$ &  $1.97$ &  $0.03$ & 0.12 & $0.01$ \\
   PG\,1559+048$^\dagger$   &  $-26.7$ &      0.9 &  8.15 &    0.06 & 6.22 &    0.12 & 0.54 &     0.03 & $217.0$ &  $2.3$ &  $27.9$ &  $1.6$ &  $18.1$ &  $1.3$ &  $1.75$ &  $0.02$ & 0.13 & $0.01$ \\
 CPD\,$-$20\,1123           &   $-6.3$ &      1.2 &  9.71 &    0.12 & 7.54 &    0.09 & 0.17 &     0.01 & $238.9$ &  $2.2$ & $-41.4$ &  $1.5$ &   $8.4$ &  $0.7$ &  $2.10$ &  $0.02$ & 0.13 & $0.01$ \\
        SB\,705             &    $4.0$ &     12.0 & 10.46 &    0.27 & 7.95 &    0.11 & 0.99 &     0.09 & $255.5$ &  $3.8$ &  $40.1$ &  $2.5$ &   $5.7$ & $11.4$ &  $2.20$ &  $0.03$ & 0.14 & $0.01$ \\
         JL\,87             &   $-6.1$ &      2.3 & 10.24 &    0.20 & 7.76 &    0.06 & 0.56 &     0.02 & $266.7$ &  $2.5$ & $-27.6$ &  $1.9$ &   $5.7$ &  $1.5$ &  $2.17$ &  $0.02$ & 0.14 & $0.01$ \\
   UVO\,0825+15$^\dagger$   &   $56.4$ &      0.5 &  9.63 &    0.10 & 7.18 &    0.10 & 0.16 &     0.01 & $232.0$ &  $2.1$ &  $46.8$ &  $1.3$ &   $6.4$ &  $0.8$ &  $2.04$ &  $0.02$ & 0.15 & $0.01$ \\
   PG\,0229+064             &    $7.6$ &      4.0 &  9.16 &    0.08 & 6.67 &    0.14 & 0.67 &     0.05 & $210.8$ &  $2.6$ &  $31.1$ &  $3.0$ &  $19.1$ &  $3.1$ &  $1.90$ &  $0.02$ & 0.16 & $0.02$ \\
  HE\,2357$-$3940           &  $-18.4$ &     14.2 &  9.59 &    0.21 & 6.95 &    0.12 & 0.96 &     0.12 & $236.1$ &  $2.6$ &  $54.5$ &  $5.1$ &   $9.6$ & $13.4$ &  $1.97$ &  $0.03$ & 0.16 & $0.02$ \\
       TON\,414             &    $2.7$ &      0.5 & 10.08 &    0.12 & 6.75 &    0.16 & 1.44 &     0.13 & $207.8$ &  $3.6$ & $-54.1$ &  $3.0$ &  $35.5$ &  $1.8$ &  $1.96$ &  $0.03$ & 0.20 & $0.02$ \\
  HE\,1310$-$2733           &   $41.5$ &      1.9 &  9.12 &    0.09 & 5.95 &    0.12 & 0.94 &     0.05 & $221.4$ &  $2.4$ & $-67.9$ &  $2.7$ &  $19.2$ &  $1.4$ &  $1.77$ &  $0.03$ & 0.21 & $0.01$ \\
  FBS\,1749+373$^\dagger$   &  $-73.6$ &      0.2 &  8.34 &    0.05 & 4.74 &    0.09 & 0.39 &     0.01 & $182.6$ &  $2.1$ &  $39.2$ &  $1.3$ &  $-2.5$ &  $0.8$ &  $1.51$ &  $0.02$ & 0.28 & $0.01$ \\
  HE\,1136$-$2504           &   $59.4$ &      9.3 &  8.40 &    0.05 & 4.63 &    0.33 & 0.68 &     0.04 & $176.0$ &  $7.8$ & $-30.4$ &  $1.6$ &   $5.6$ &  $5.5$ &  $1.48$ &  $0.07$ & 0.29 & $0.04$ \\
  TYC\,3519-907-1           &  $-62.7$ &      0.2 &  9.30 &    0.08 & 5.12 &    0.09 & 0.30 &     0.01 & $195.5$ &  $2.1$ &  $73.5$ &  $2.2$ &   $2.7$ &  $0.9$ &  $1.66$ &  $0.02$ & 0.29 & $0.01$ \\
  PG\,0240+046              &   $63.4$ &      2.0 & 10.57 &    0.11 & 5.36 &    0.13 & 0.57 &     0.03 & $196.9$ &  $3.1$ &  $93.4$ &  $2.8$ &   $0.6$ &  $2.1$ &  $1.78$ &  $0.03$ & 0.33 & $0.02$ \\
         HZ\,44$^\dagger$   &  $-12.7$ &      0.4 &  9.30 &    0.07 & 4.48 &    0.11 & 0.41 &     0.02 & $177.3$ &  $2.7$ &  $76.8$ &  $2.3$ &  $10.2$ &  $0.8$ &  $1.52$ &  $0.02$ & 0.35 & $0.02$ \\
 EC\,22133$-$6446           &  $-21.0$ &     10.0 &  9.30 &    0.20 & 3.58 &    0.26 & 0.44 &     0.05 & $159.1$ &  $7.9$ & $101.3$ &  $8.7$ &  $-2.9$ &  $7.2$ &  $1.31$ &  $0.07$ & 0.44 & $0.04$ \\
 EC\,22536$-$5304$^\dagger$ &   $-3.3$ &      5.0 &  8.21 &    0.05 & 2.51 &    0.13 & 1.50 &     0.12 & $113.2$ &  $4.2$ & $-26.4$ &  $2.8$ &  $58.3$ &  $4.3$ &  $0.93$ &  $0.04$ & 0.53 & $0.02$ \\
      Feige\,46$^\dagger$   &   $90.0$ &      4.0 &  9.64 &    0.09 & 1.13 &    0.12 & 0.68 &     0.04 &  $63.5$ &  $5.2$ & $115.5$ &  $3.7$ &  $13.8$ &  $4.4$ &  $0.55$ &  $0.04$ & 0.79 & $0.03$ \\
 \lsiv$^\dagger$            & $-154.0$ &      1.0 &  8.15 &    0.05 & 0.66 &    0.14 & 0.24 &     0.02 & $-42.5$ &  $6.8$ &  $21.2$ &  $2.7$ &   $0.1$ &  $2.9$ & $-0.35$ &  $0.06$ & 0.85 & $0.03$ \\
\bottomrule
\end{tabular}
\end{center}
\end{sidewaystable*} %

\section{Full spectral comparisons}

This section presents the full spectral comparison between the observed spectra of \EC\ and our final synthetic spectra.
We show the co-added 2019 HRS spectra and two examples for the UVES spectra. 
The 2019 HRS spectra were shifted to a common radial velocity for the \up{sdOB} component before they were co-added. 
Lines that originate from the \up{sdF} component are therefore \up{very} slightly broadened in this spectrum. 
The observations were normalised to match the continuum levels of our synthetic spectra.
Telluric lines in the red spectra were removed using the grid of transmission spectra provided by \cite{Moehler2014}.
The strongest metal lines for the sdOB are labelled on top, while lines of the sdF are labelled at the bottom.
The combined model is shown in red, while the contributions of the sdOB and sdF components are shown in blue and dark red, respectively. 
\newpage

\captionsetup[ContinuedFloat]{labelformat=continued}
\begin{figure*}%
\centering
\includegraphics[width=24.2cm,angle=90,page=1]{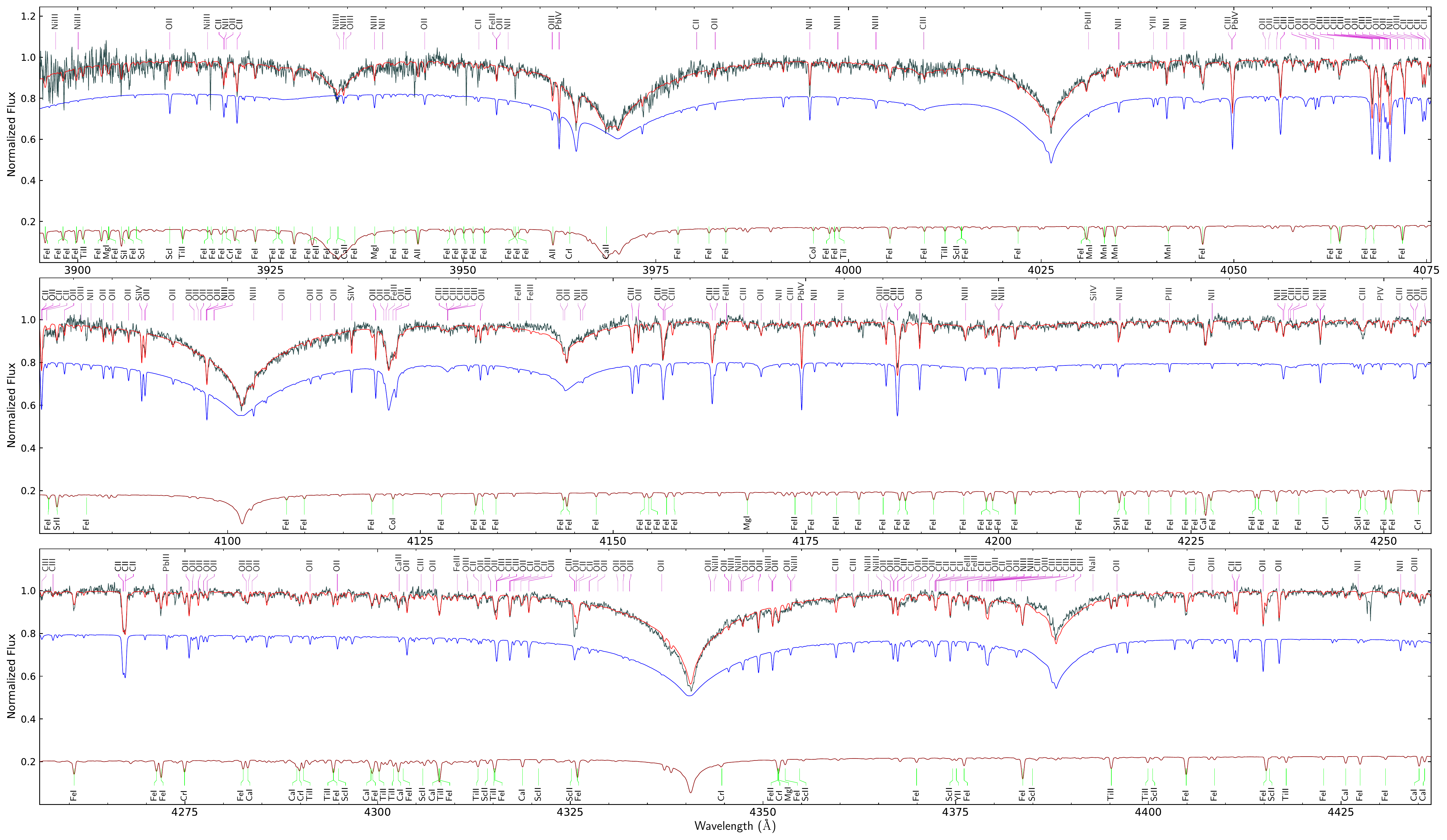}
\caption{HRS spectrum of \EC\ (grey) and the combined (red) and individual models.}
\label{fig:hrs:full}
\end{figure*}
\begin{figure*}%
\ContinuedFloat
\centering
\includegraphics[width=24.2cm,angle=90,page=2]{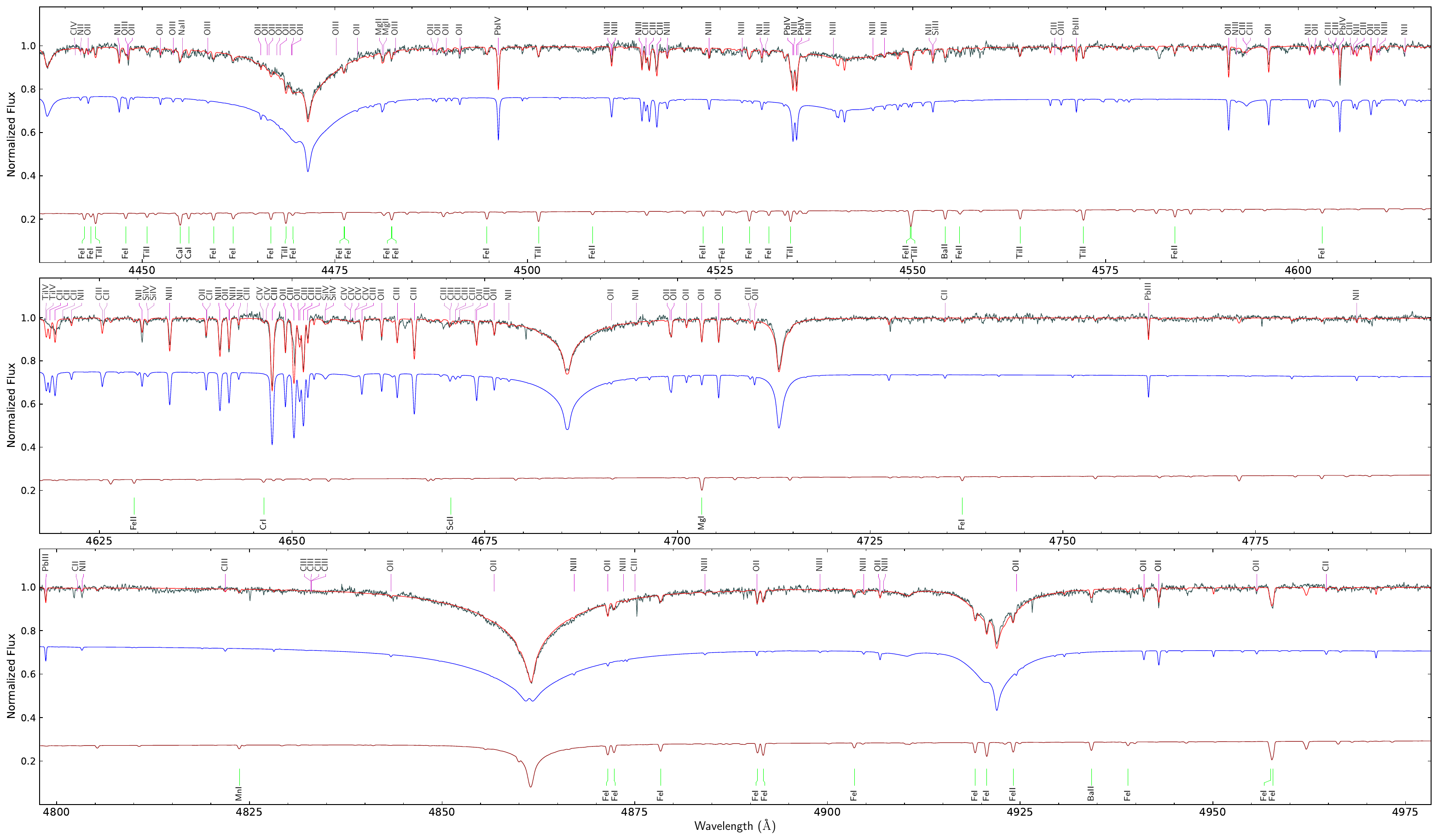}
\caption{Blue HRS spectrum of \EC\ (grey) and the combined (red) and individual models.}
\end{figure*}
\begin{figure*}%
\ContinuedFloat
\centering
\includegraphics[width=24.2cm,angle=90,page=3]{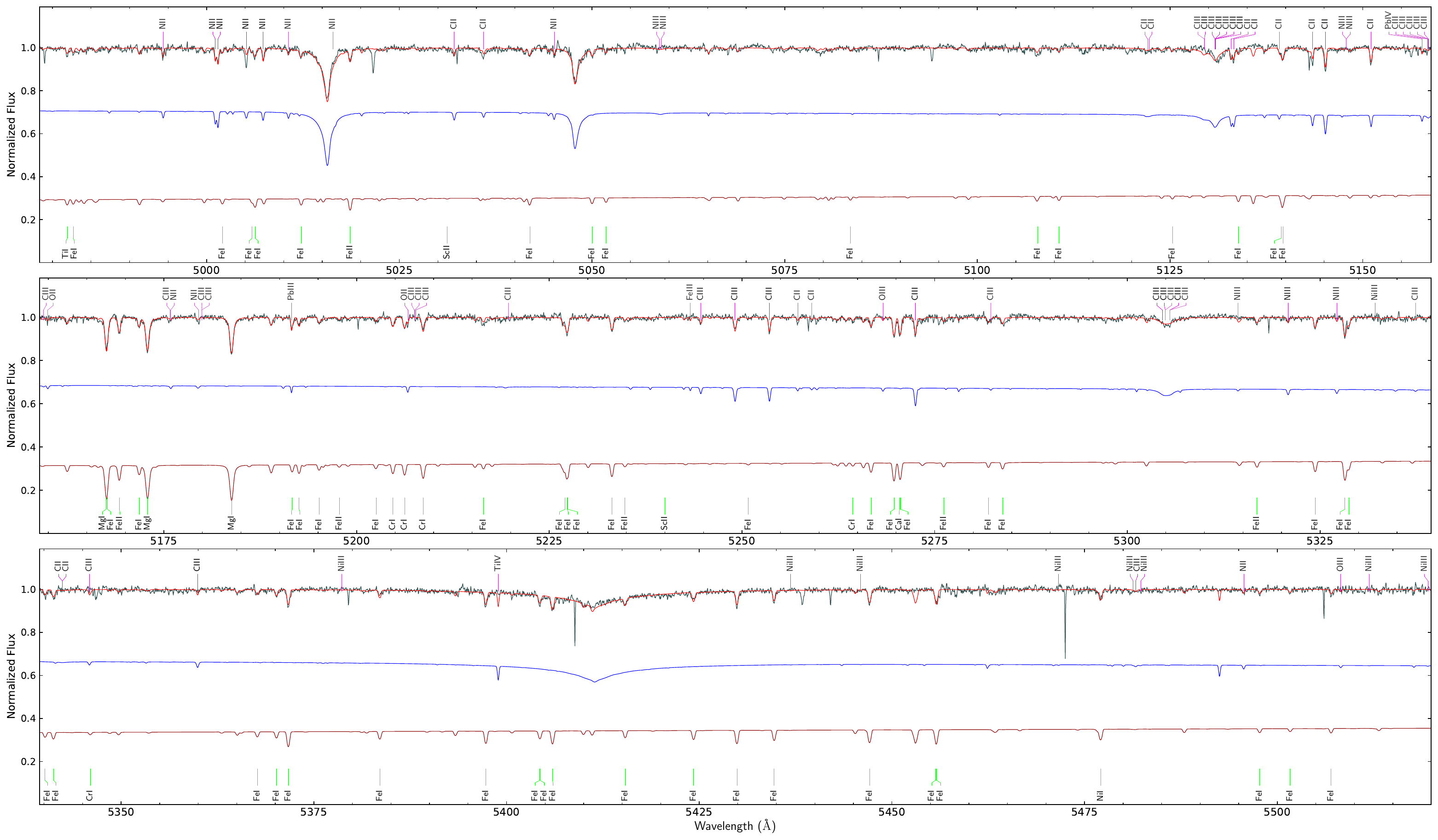}
\caption{Blue HRS spectrum of \EC\ (grey) and the combined (red) and individual models.}
\end{figure*}

\captionsetup[ContinuedFloat]{labelformat=continued}
\begin{figure*}%
\centering
\includegraphics[width=24.2cm,angle=90,page=1]{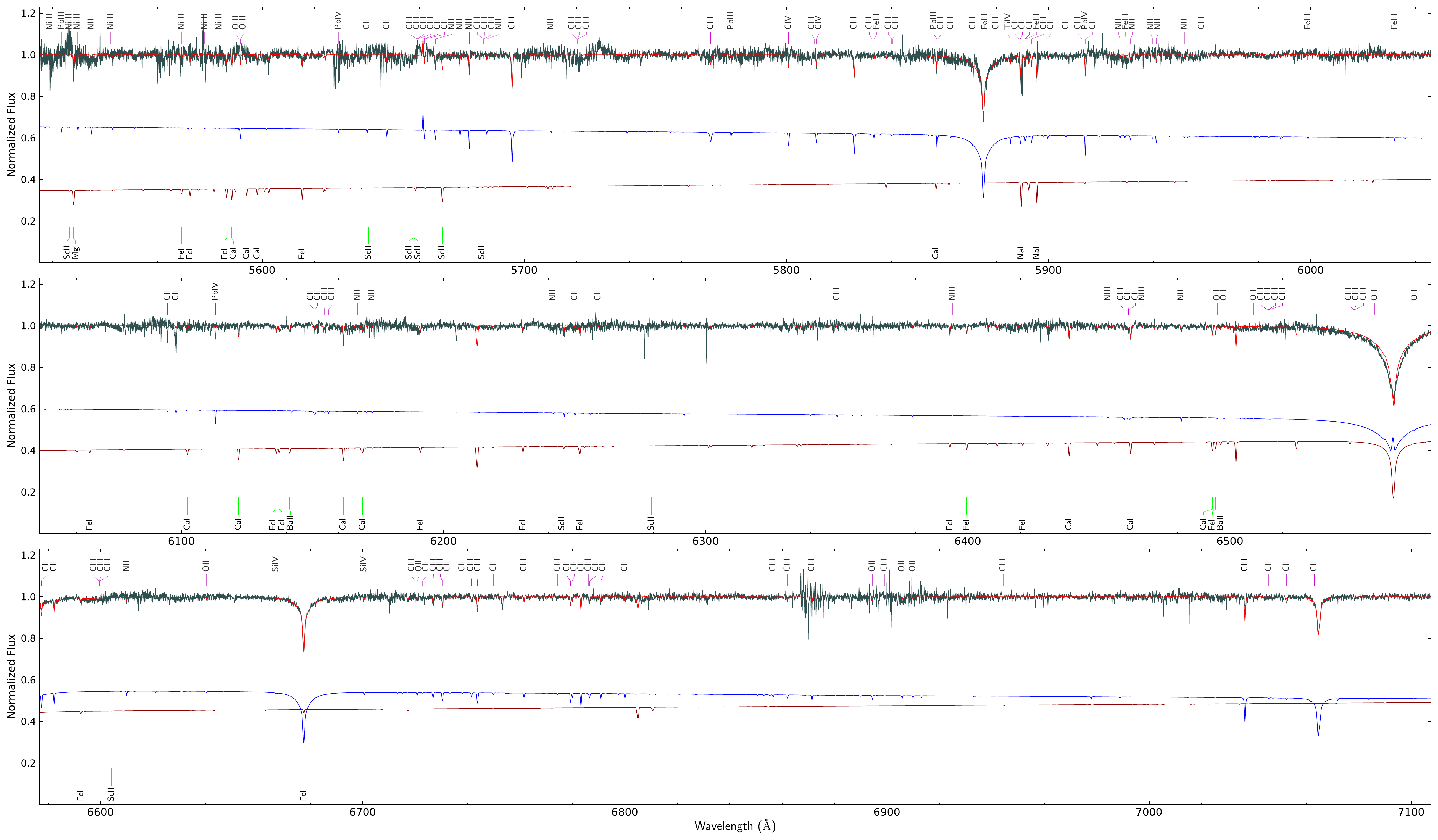}
\caption{Red HRS spectrum of \EC\ (grey) and the combined (red) and individual models.}
\end{figure*}
\begin{figure*}%
\ContinuedFloat
\centering
\includegraphics[width=24.2cm,angle=90,page=2]{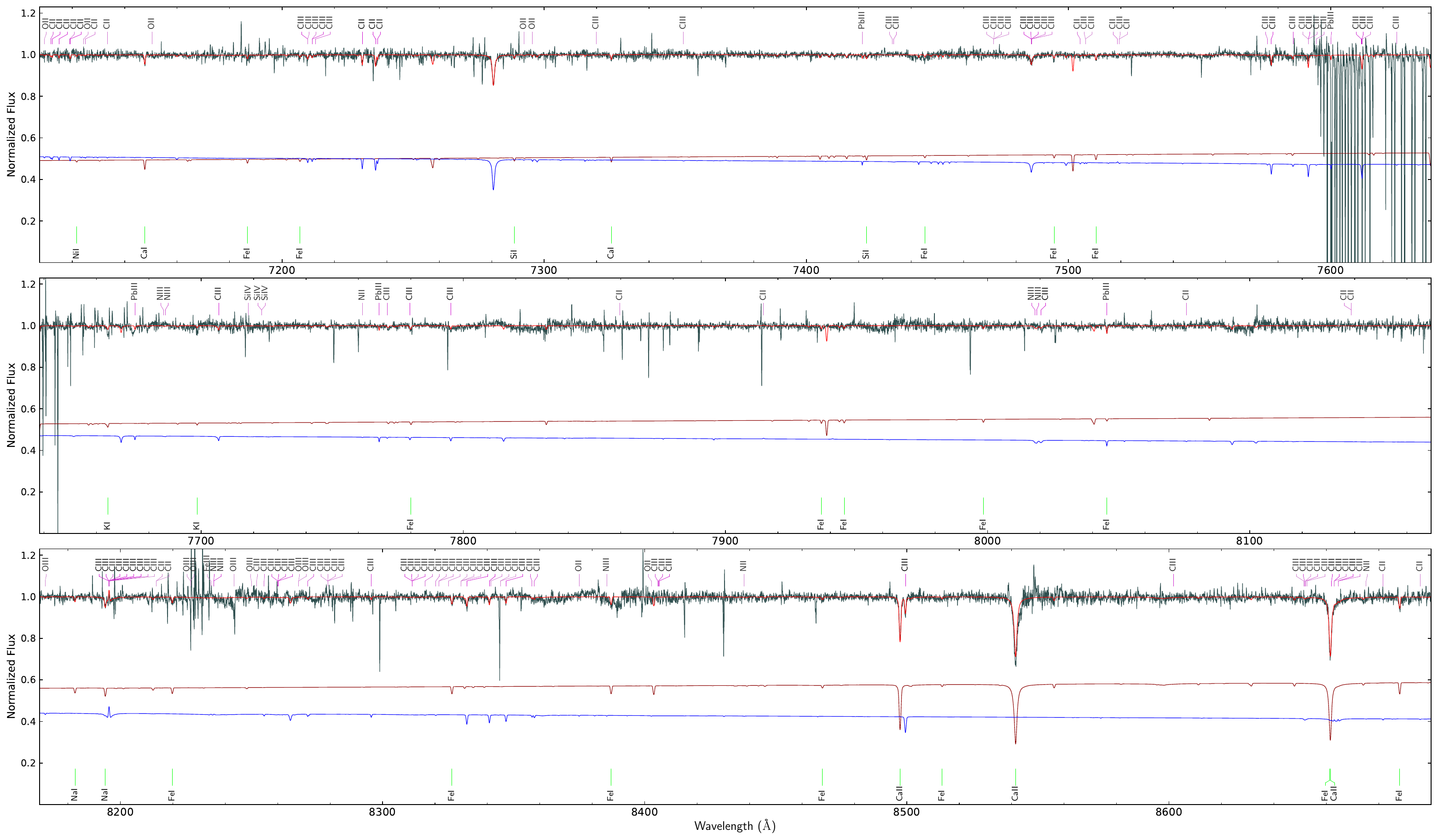}
\caption{Red HRS spectrum of \EC\ (grey) and the combined (red) and individual models.}
\end{figure*}

\captionsetup[ContinuedFloat]{labelformat=continued}
\begin{figure*}%
\centering
\includegraphics[width=24.2cm,angle=90,page=1]{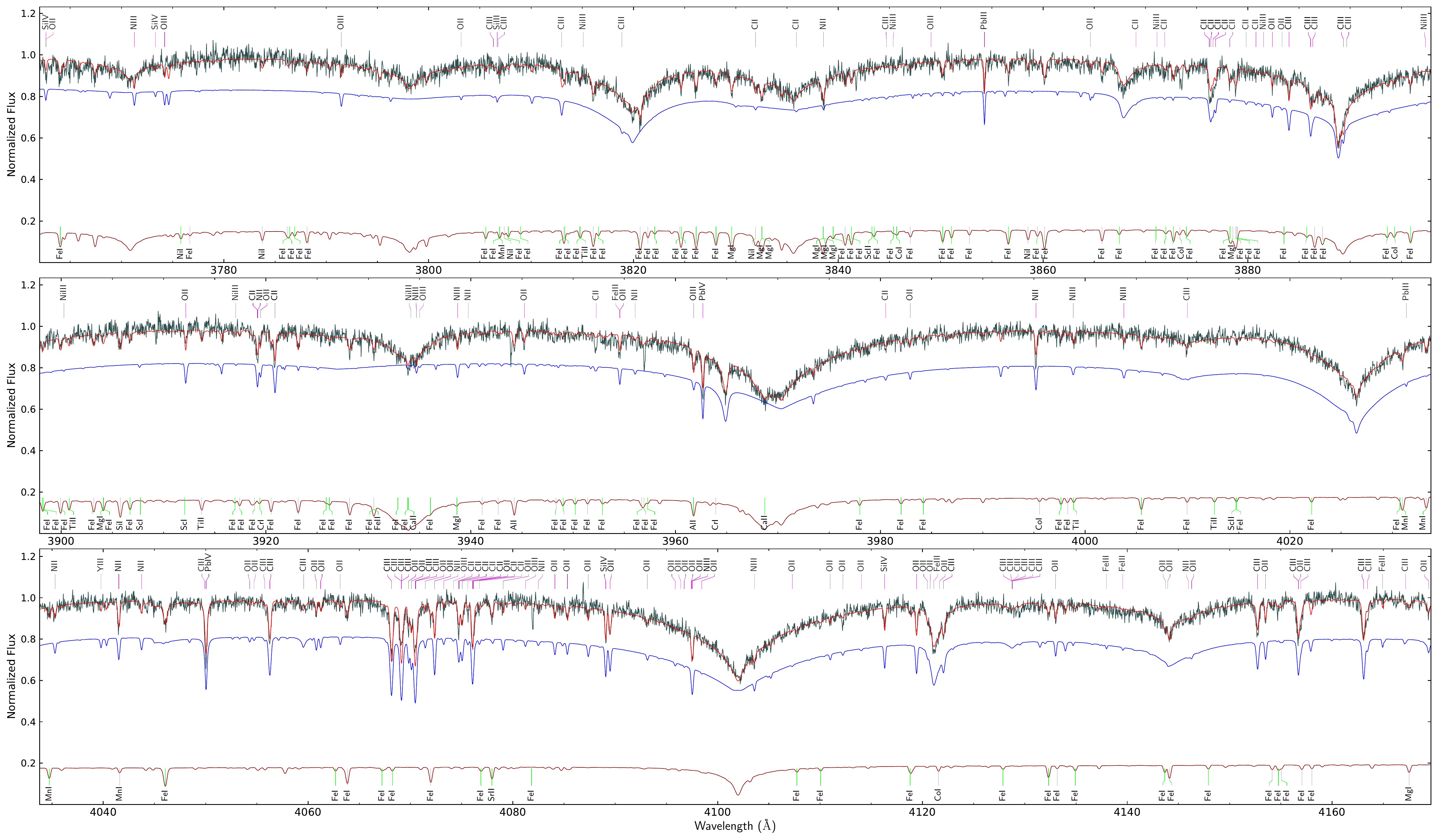}
\caption{Blue UVES spectrum of \EC\ (grey) and the combined (red) and individual models.}
\end{figure*}
\begin{figure*}%
\ContinuedFloat
\centering
\includegraphics[width=24.2cm,angle=90,page=2]{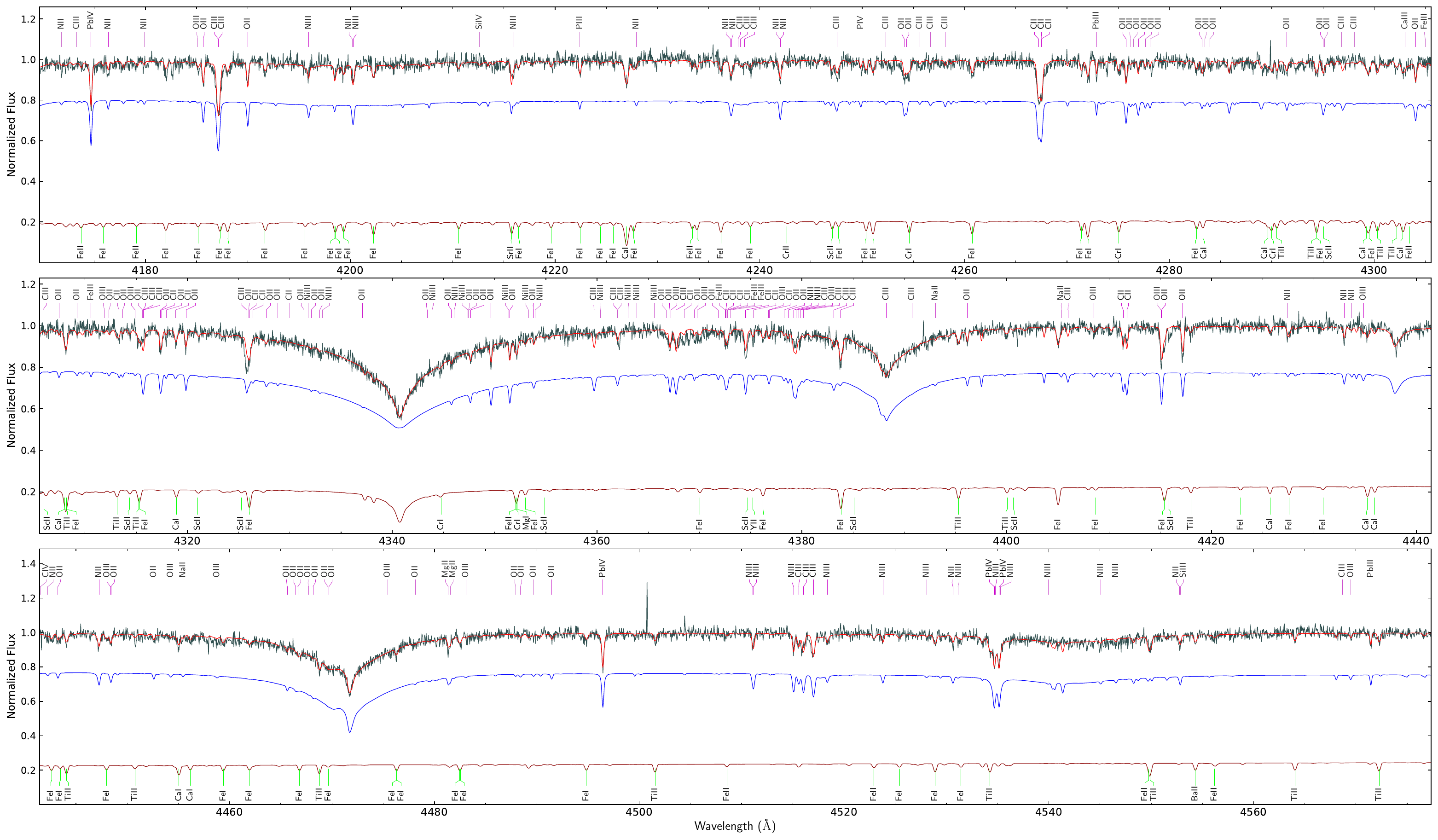}
\caption{Blue UVES spectrum of \EC\ (grey) and the combined (red) and individual models.}
\end{figure*}
\begin{figure*}%
\ContinuedFloat
\centering
\includegraphics[width=24.2cm,angle=90,page=3]{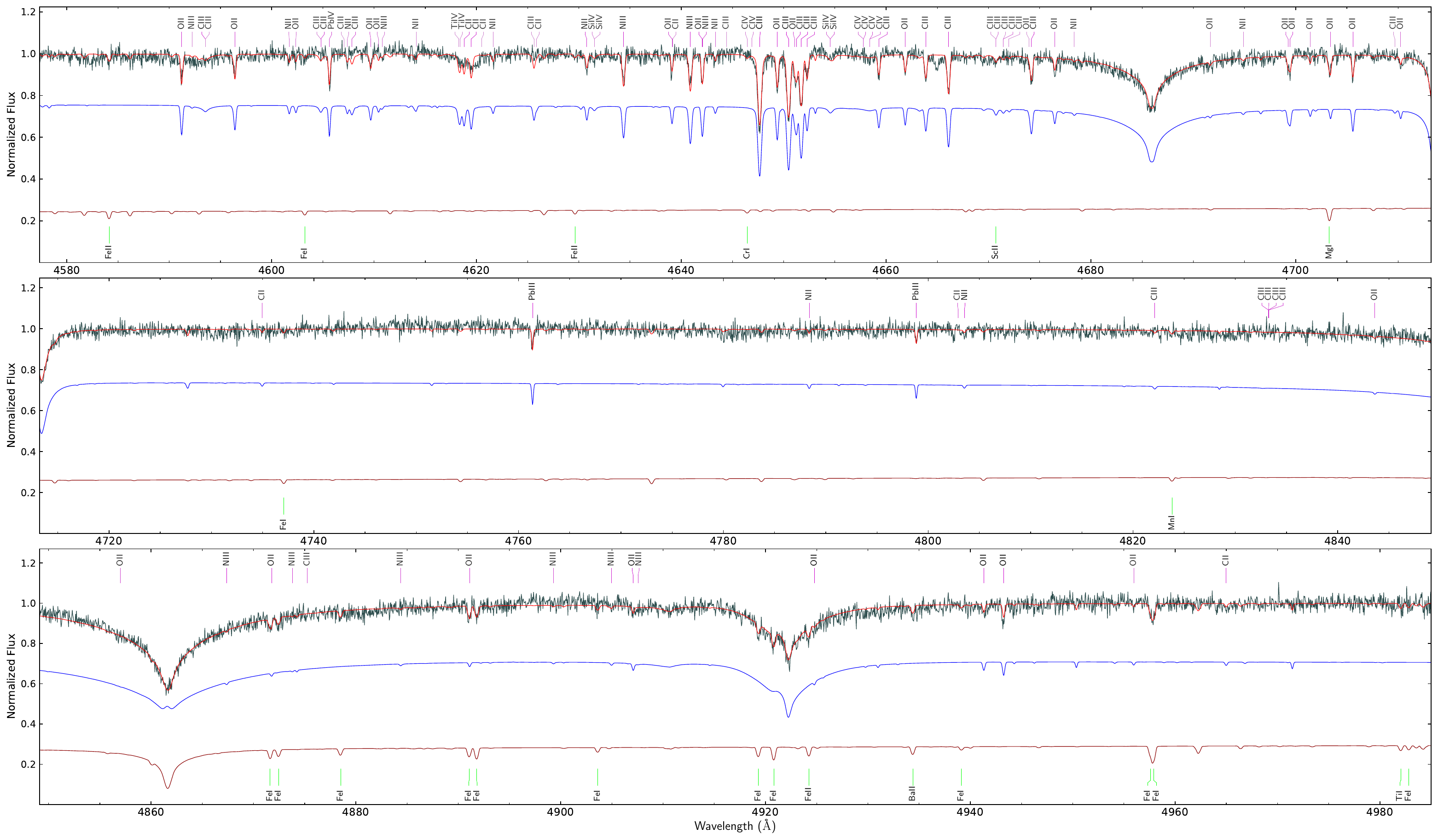}
\caption{Blue UVES spectrum of \EC\ (grey) and the combined (red) and individual models.}
\end{figure*}

\captionsetup[ContinuedFloat]{labelformat=continued}
\begin{figure*}%
\centering
\includegraphics[width=24.2cm,angle=90,page=1]{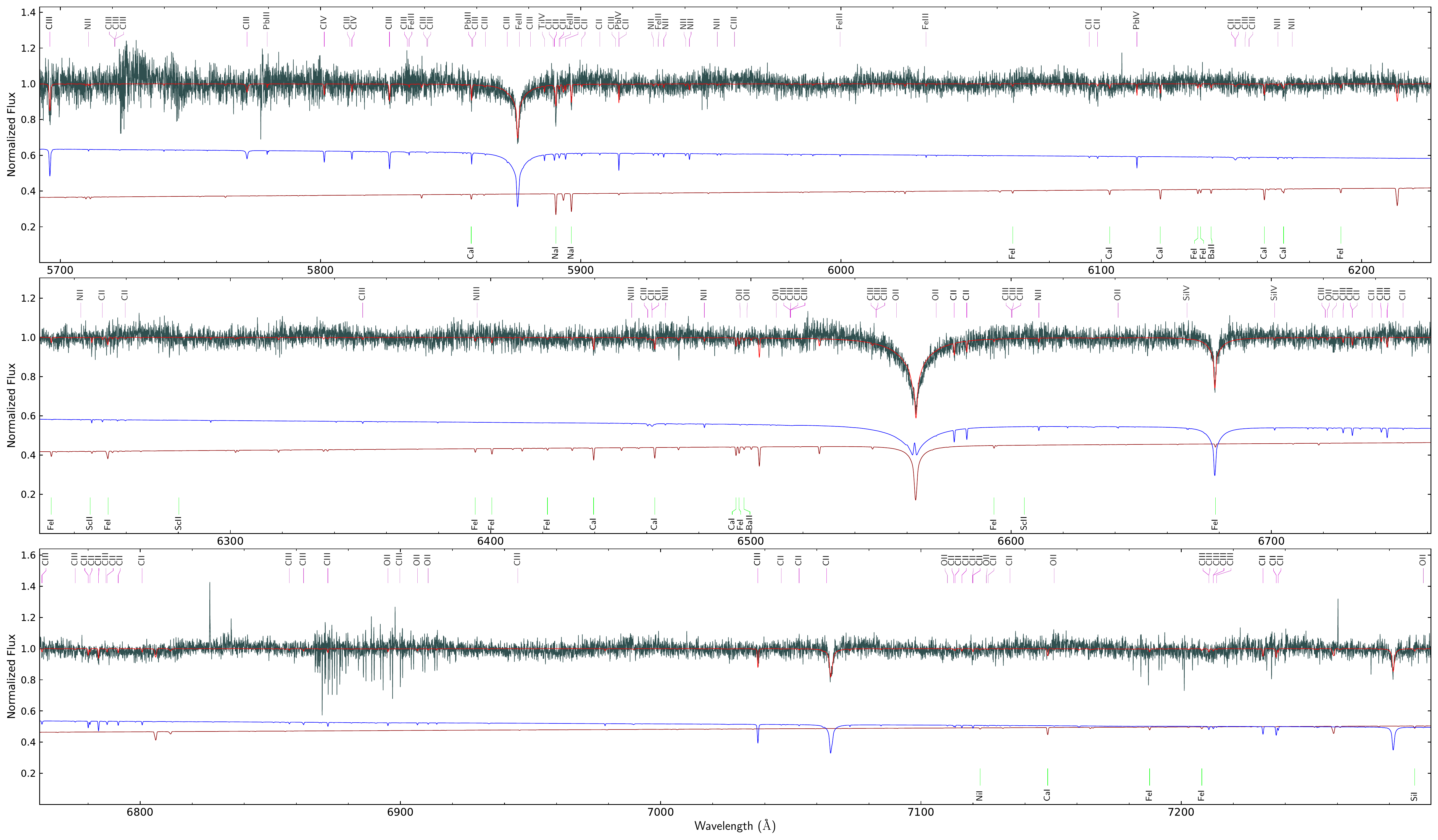}
\caption{Red UVES spectrum of \EC\ (grey) and the combined (red) and individual models.}
\end{figure*}
\begin{figure*}%
\ContinuedFloat
\centering
\includegraphics[width=24.2cm,angle=90,page=2]{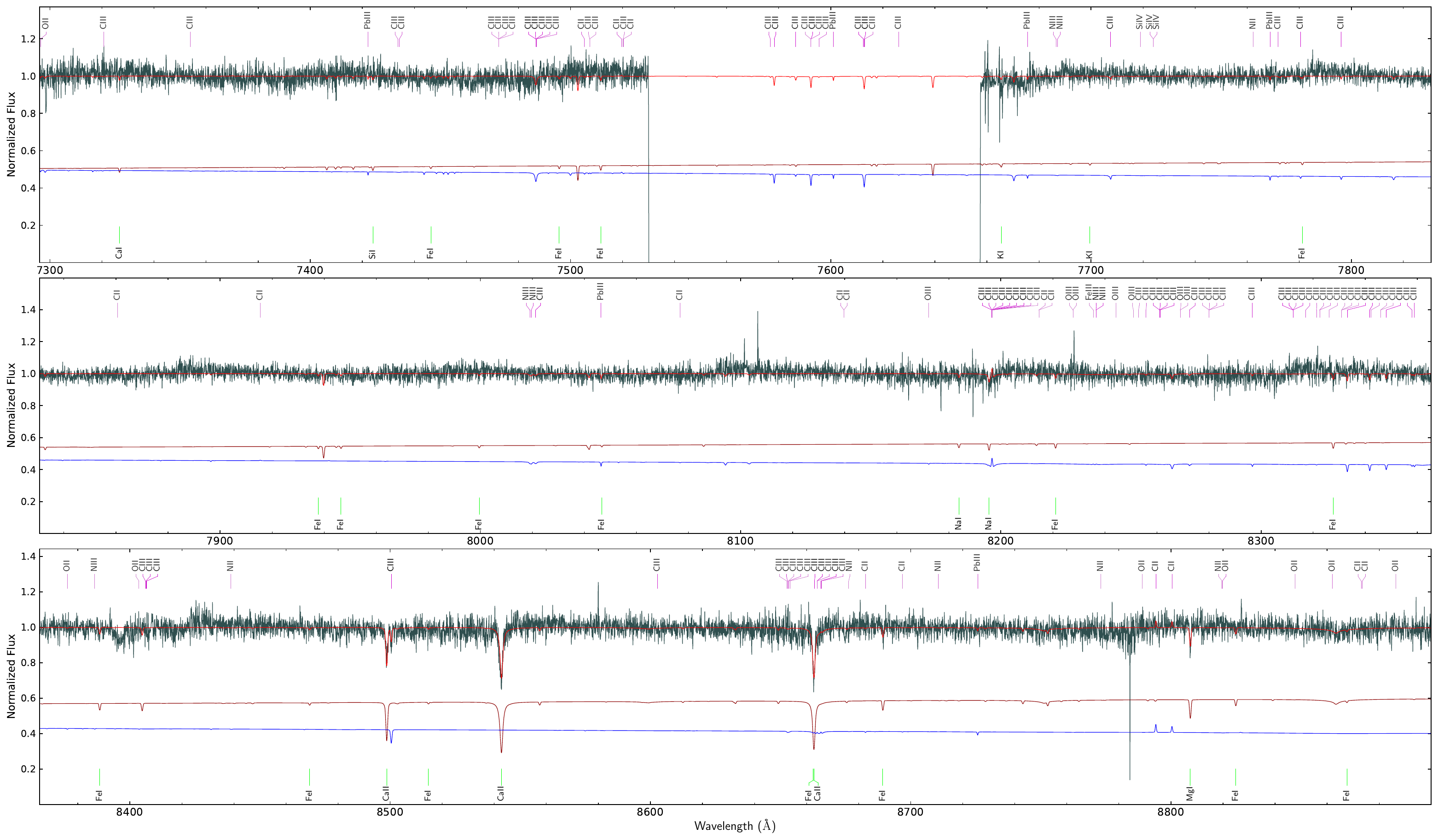}
\caption{Red UVES spectrum of \EC\ (grey) and the combined (red) and individual models.}
\end{figure*}

\end{appendix}

\end{document}